\documentclass[usenatbib]{mn2e}
\usepackage{natbibmnfix,graphicx,times,array,ctable}

\newcommand{\Msun}{\mathrm{M}_{\odot}}
\newcommand{\rcut}{r_{\mathrm{cut}}}
\newcommand{\rcore}{r_{\mathrm{core}}}
\newcommand{\kpc}{\mathrm{kpc}}
\newcommand{\Mpc}{\mathrm{Mpc}}
\newcommand{\PA}{\mathrm{PA}}

\title[Cosmography with cluster strong lenses]{Cosmography with cluster strong lenses: the influence of substructure and line-of-sight halos}

\author[D'Aloisio \& Natarajan]{Anson
  D'Aloisio$^1$\thanks{Email: anson.daloisio@yale.edu} \& 
	Priyamvada Natarajan$^{1,2}$\thanks{Email: priya.natarajan@yale.edu}\\
$^1$Department of Physics, Yale University, PO Box 208120, New
    Haven, CT 06520-8120\\
$^2$Department of Astronomy, Yale University, PO Box 208101, New Haven, CT 06511}

\begin{document}

\maketitle

\begin{abstract}
We explore the use of strong lensing by galaxy clusters to constrain
the dark energy equation of state and its possible time
variation. The cores of massive clusters often contain several multiply imaged systems of background galaxies at different
redshifts.  The locations of lensed images can be used to constrain cosmological parameters due to their dependence on the ratio of angular diameter distances. We employ Monte-Carlo
simulations of cluster lenses, including the contribution from
substructures, to assess the feasibility of this potentially powerful
technique.  At the present, parametric lens models use well motivated scaling relations between mass and light to incorporate cluster member galaxies, and do not explicitly model line-of-sight structure.  Here, we quantify modeling errors due
to scatter in the cluster galaxy scaling relations and un-modeled line-of-sight halos.  These errors are of the
order of a few arcseconds on average for clusters located at typical
redshifts ($z \sim 0.2 -0.3$).  Using Bayesian Markov Chain
Monte-Carlo techniques, we show that the inclusion of these modeling
errors is critical to deriving unbiased constraints on dark energy.
However, when the uncertainties are properly quantified, we show that
constraints competitive with other methods may be obtained by
combining results from a sample of just $10$ simulated clusters with
$20$ families each.  Cosmography with a set of well studied cluster
lenses may provide a powerful complementary probe of the dark energy
equation of state. Our simulations provide a convenient method of
quantifying modeling errors and assessing future strong lensing survey
strategies.
\end{abstract}


\section{INTRODUCTION}

The recent influx of cosmological data from a variety of complementary
observations has led to the development of a highly successful
concordance cosmology.  The overall picture that has emerged, where
the most significant contributions to the Universe's energy content
are baryonic matter ($\sim 5$ per cent), non-relativistic dark matter
($\sim 20$ per cent), and dark energy ($\sim$ 75 per cent), appears to
be consistent with all current observations.  A form of dark energy is
required within the Friedmann-Robertson-Walker framework to explain
the current epoch of accelerated expansion, which is most directly
evident in the Hubble diagram of Type Ia supernovae
\citep{1998AJ116.1009R,1999ApJ...517..565P}.  However, a variety of
other combined probes such as the Wilkinson Anisotropy Probe (WMAP)
\citep[e.g.][]{2010arXiv1001.4538K}, baryon acoustic oscillations
\citep{2002MNRAS.330L..29E,2005PhRvD..71j3515S,2005ApJ...633..560E}
cluster abundances \citep{2009ApJ...692.1060V}, cosmic shear
measurements
\citep{2000MNRAS.318..625B,2000astro.ph..3338K,2000A&A...358...30V,2000Natur.405..143W,2006A&A...452...51S},
and cluster baryon fractions \citep{2004MNRAS.353..457A} also provide
compelling evidence for its existence.

Dark energy is typically parameterized by an equation-of-state with
the form $P = w_x \rho$, where $w_x < -1/3$ in the current epoch. Time
variation or observed deviations from $w_x=-1$ at the present day
could yield important clues to its nature.  In the last decade, much
effort has been devoted to further constraining the equation of state.
Owing to the influence of systematic errors in astrophysical
measurements, the development of complementary observational
techniques is vital to this task.  In this spirit, we investigate the
use of cluster strong lensing (CSL) as a complementary technique to
constrain $w_x$.

The possibility of constraining cosmology with CSL systems has been
 explored in the past
 \citep[e.g.][]{1981ApJ...248L.101P,1998ApJ...502...63L,1999ApJ...524..504C,2002A&A...387..788G,2002A&A...393..757S,2004A&A...417L..33S,2005ApJ...622...99D,Meneghetti:2005ax,2005A&A...442..413M,2005MNRAS.361.1250M,2009MNRAS.396..354G,Jullo2010Sci}.
 The abundance of arcs may provide useful cosmological
 constraints. \citet{2005A&A...442..413M} and
 \citet{2005MNRAS.361.1250M} explored the statistics of arcs in
 various dark energy cosmologies.  They found that the variation in
 arc abundances, particularly at higher redshifts, can potentially be
 used to differentiate between dark energy models. 

The locations of images in CSL systems also contain useful
 cosmological information.  These image positions depend not only on
 the mass distribution, but also on the angular diameter distances
 between the observer, lens, and source.  If more than one set of
 images is observed, the geometrical dependence may be exploited to
 probe the cosmological parameters.  In this paper we will explore the
 latter technique in greater detail.
 
Since the cores of clusters have surface densities which are typically
much larger than the critical surface density for multiple image
production, they are prime locations for identifying strongly lensed
images. Indeed, cluster lenses containing an abundance of multiple
images have already been utilized to place tight constraints on mass
distributions in the inner regions of clusters. The most well studied
CSL system to date is Abell 1689
\citep{2005ApJ...621...53B,2006MNRAS.372.1425H,Limousin:2007cq,Jullo2010Sci}.  At present, strongly lensed images from 42 unique sources have been
identified.  Utilizing the large number of constraints and high
resolution of the Hubble Space Telescope (HST) Advanced Camera for
Surveys (ACS), the parametric mass models cited above can reproduce
the observed image locations to within a few arcseconds.  Similar,
though less detailed, strong lensing analyses have been carried out
for a growing sample of clusters
\citep[e.g.][]{2002ApJ...580L..11N,2003ApJ...598..804K,2007arXiv0710.5636E,2008A&A...489...23L,
2009ApJ...693..970N,2009A&A...498...37R,2009ApJ...707L.163S,
2010MNRAS.402L..44R, 2010MNRAS.404..325R}.

Preliminary work has been done on bringing these rich systems to bear
on cosmological parameters.  \citet{2002A&A...387..788G} explored
simulated constraints on the mean matter density $\Omega_m$ and $w_x$
using a single cluster with three strongly lensed sources.
\citet{2002A&A...393..757S} applied this technique to the cluster Cl0024+1654 at $z = 0.4$. Their results were
found to be consistent with a flat, accelerating universe.
\citet{2004A&A...417L..33S} performed a more detailed application to
Abell 2218 using only $4$ multiple-image systems.  Assuming a flat
universe, their results are consistent with $\Omega_m < 0.3$ and $w_x
< -0.85$.  Most recently, \citet{Jullo2010Sci} obtained $\Omega_m = 0.25 \pm 0.05$ and $w_x = -0.97\pm0.07$ by combining a strong lensing analysis of Abell 1689 with WMAP and X-Ray cluster constraints.

Theoretical investigations of this technique
using simulations were performed by \cite{2005ApJ...622...99D}.  They ray-traced through N-body simulations to create mock strong lensing image catalogs.  They
then attempted to recover the input cosmology by fitting a parametric
NFW model and found that significant biases can result.  Owing to the
relatively large errors due to complexities in the mass distribution
and line-of-sight (LOS) structure, they conclude that a single lensing
system may not provide significant leverage on cosmological
parameters.

Perhaps the most obvious way to overcome these difficulties is to
obtain a larger sample of CSL systems.  This approach has the
advantage that results obtained from different lines of sight are
statistically independent.  Hence, their results may be combined in a
trivial manner.  More recently, \citet{2009MNRAS.396..354G} explored
the prospect of {\bf combining} CSL systems as a more powerful probe
of dark energy.  They found that competitive constraints can be
obtained by combining at least $10$ lenses with $5$ or more image
systems.  Note that space-based measurements, with positional errors
$\sim 0.1$ arcseconds, and spectroscopic redshift determinations
($\sigma_z \sim 0.001$) are a necessity.  The current work expands
upon the \citet{2009MNRAS.396..354G} investigation.  We further
explore the issue of overcoming large modeling uncertainties by
combining a sample of CSL systems.  We utilize more realistic
simulations of cluster lensing systems that include sub-structure in
the lens plane and line-of-sight structure to address several sources
of modeling errors.  Moreover, instead of a maximum likelihood
routine, a Bayesian Markov Chain Monte-Carlo (MCMC) technique is used
to probe the full parameter space, allowing us to marginalize over all
lens parameters and draw more robust conclusions on the ability of CSL
to constrain cosmological parameters.
 
This paper is organized as follows.  In Section
\ref{SEC:imagelocations} we briefly discuss the sensitivity of CSL to
cosmological parameters. In Section \ref{SEC:methods} we describe our
simulations and discuss parameter recovery through Bayesian MCMC.  In
Section \ref{SEC:modelingerrors} we use our Monte-Carlo simulations to
explore modeling uncertainties arising from complexities in the
cluster galaxy population.  We also perform a simple investigation of
errors due to correlated LOS structure, and a more detailed analysis
of uncorrelated LOS halos.  In Section \ref{SEC:simulatedconstraints},
we explore the prospect of overcoming these errors using a sample of
10 simulated clusters.  We derive simulated constraints on the dark
energy equation-of-state for this sample.  In Section
\ref{SEC:clustermass}, we explore how differences in the large-scale
mass distribution of the lens, including bi-modality, affect
cosmological constraints.  Finally, we offer concluding remarks in
Section \ref{SEC:discussion}.
  
\section{Image locations and the dark energy equation-of-state}
\label{SEC:imagelocations}

In this section, we briefly outline the sensitivity of CSL to
cosmological parameters.  For a more detailed discussion, see
\citet{2002A&A...387..788G} and \cite{2009MNRAS.396..354G}.  The lens
equation is given by

\begin{equation}
\vec{\beta}_i = \vec{\theta_i} - \frac{2}{c^2}\frac{D_{ls}}{D_{ol} D_{os}} \nabla\phi(\vec{\theta_i}),
\label{lensEQ}
\end{equation}
where the angular coordinates of the source $i$ and its corresponding
image(s) are given by $\vec{\beta}_i$ and $\vec{\theta}_i$
respectively, and $\phi$ is the projected Newtonian potential of the
lens.  For the parametric models used in this work, the potential is
typically normalized by the associated central velocity dispersion,
$\sigma_v$.  The subscripts $o$, $l$, and $s$ correspond to the
observer, lens, and source.  We define $D_{ab}$ to be the angular
diameter distance from $z_a$ to $z_b$.  In the case of a flat,
two-component universe, $D_{ab}$ is given by

\begin{equation}
D(z_a,z_b) = \frac{c/H_0}{1+z_b}\int^{z_b}_{z_a}{\mathrm{d}z~\left( \Omega_m(1+z)^3 + \Omega_X(z) \right)^{-1/2}},
\end{equation}
where $H_0$ is the present day Hubble constant, $c$ is the speed of
light, and $\Omega_m$ is the present day matter density normalized by
the critical density.  The function $\Omega_X(z)$ is the contribution
from dark energy and its form depends on the choice of
parameterization.  In this work we consider two
parameterizations: 1) a constant equation of state, $w_x$. 2) the
widely used Chevallier, Polarski, and Linder (CPL) parameterization,
$w_x(z) = w_0 + w_a z/(1+z)$ \citep{Chevallier:fk,Linder:2003zm}.  In
this case, $\Omega_X(z)$ is given by

\begin{equation}
\Omega_X(z) = \Omega_X (1+z)^{3(1+w_0+w_a)} \exp\left[ -\frac{3 w_a z}{1+z} \right].
\end{equation}
Note that the angular diameter distance factor in (\ref{lensEQ})
contains all of the explicit dependence on the cosmological
parameters, while the influence of the mass distribution comes in
through the gradient term.  The CSL technique discussed in this paper
is a purely geometric probe and utilizes only the angular diameter
distances for constraints.  Aside from the abundance of image systems,
the influence of cosmology on the detailed properties and formation 
of lensing structures does not come into play.

Consider a single source at $\vec{\beta}_i$ that is lensed into a set
of $n$ multiple images $\vec{\theta}_{i,j}$, where $j = 1,...,n$.  In
this case, it is impossible to isolate the effect of the cosmological
parameters because the geometric factor in equation (\ref{lensEQ}) is
completely degenerate with the normalization of the projected lens
potential.  If more than one source is lensed into multiple image
families, as is typically the case in CSL systems, then the above
degeneracy is broken.  Consider the case of $N$ sources at different
redshifts.  In this case, there is a set of $N$ lens equations, each
given by
\begin{equation}
\begin{array}{ll}  \vec{\beta}_1 =&  \vec{\theta_1} - \frac{2}{c^2}\Gamma^{(1)} \nabla\phi(\vec{\theta_1}) \\ ~\vdots & ~\vdots \\
	 \vec{\beta}_N =&  \vec{\theta_N} - \frac{2}{c^2}\Gamma^{(N)} \nabla\phi(\vec{\theta_N}), \end{array}
\label{manylensEQ}
\end{equation}
where the $\Gamma$'s represent the geometric factors.  The
superscripts on $\Gamma$ denote the different source redshifts.  Note
that equation (\ref{manylensEQ}) applies to a single cluster lens so
that the normalization of $\phi$ is the same for each line.  Owing to
the $N$ different source redshifts involved, a change of the
cosmological parameters may now be discerned from changes in
$\sigma_v$.  Hence the dark energy equation-of-state may be probed
using a set of multiply imaged systems as constraints.

\begin{figure}
\begin{center}
\resizebox{8.0cm}{!}{\includegraphics{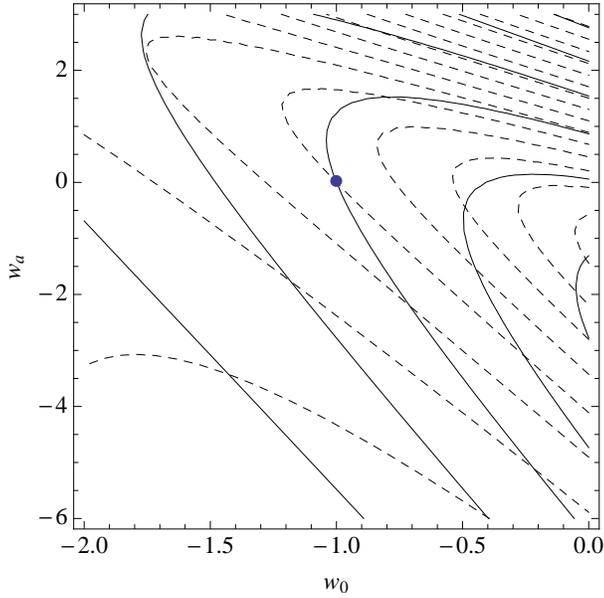}}
\end{center}
\caption{Contours of the ratio $\Gamma^{(1)}/\Gamma^{(2)}$ in the
$w_0-w_a$ plane for lensing configurations with two multiply-imaged
families from background galaxies at 2 distinct redshifts. The solid 
(dashed) curves correspond to $z_l = 0.2(0.3)$, $z_{s1} = 0.7(1.0)$, 
and $z_{s2} = 1.5(4)$.}
 \label{FIG:CFR}
\end{figure}

It is instructive to consider the simple example of a lens at $z_l$
and two families at $z_{s1}$ and $z_{s2}$.  In this case, the two lens
equations may be combined by eliminating the normalization of $\phi$.
The effective dependence on cosmology for the system is through the
ratio $\Gamma^{(1)}/\Gamma^{(2)}$.  Figure~\ref{FIG:CFR} shows
contours of $\Gamma^{(1)}/\Gamma^{(2)}$ in the $w_0-w_a$ plane (CPL model) 
for two different lensing configurations \citep[for the constant $w_x$ case,
see][]{2009MNRAS.396..354G}. The solid (dashed) curves correspond to
$z_l = 0.2 (0.3)$, $z_{s1} = 0.7(1.0)$ and $z_{s2} =1.5(4)$.  The
contours, which correspond to one per cent differences in the
geometric ratios, illustrate degeneracies in the $w_0-w_a$ plane for
CSL cosmography with two families.  Degeneracies may be broken by
using a variety of lensing configurations with different lens and
source redshifts.

\section{Methods}
\label{SEC:methods}

\subsection{Simulations of Cluster Strong Lenses}
\label{SEC:simulations}
 
In this section, we describe how our Monte-Carlo simulations of
cluster lenses are generated.  In the bulk of this work, we employ
simulated clusters with two components: 1) a smooth large-scale
density profile.  2) galaxy sized sub-structures within the core of
the cluster.  Lenses with more complicated large-scale components are
known to exist \citep[for example,
see][]{Limousin:2007cq,2009MNRAS.395.1319J}.  We also address cluster
bi-modality in Section \ref{SEC:clustermass}.

We model all cluster and galaxy scale halos with a smoothly truncated
version of the Pseudo-Isothermal Elliptical Mass Distribution (PIEMD)
in \citet{Kassiola:1993la} \citep[see][]{1996ApJ...471..643K}.  The
profile is characterized by a central velocity dispersion $\sigma_v$,
core radius $\rcore$, scale radius $\rcut$, and ellipticity parameter
$\epsilon$.  In addition, the profile center $( x_0,y_0)$ and position
angle $\theta_{\mathrm{PA}}$ are used to characterize lens the
orientation.  For the circular case, $\rcore$ marks the transition
from a constant density to $\rho \sim r^{-2}$.  Outside of $\rcut$,
the density drops off rapidly as $\rho \sim r^{-4}$.  The truncated
PIEMD profile has been used to successfully model both cluster and
galaxy scale lenses in the past
\citep[e.g.,][]{1998ApJ...499..600N,2002ApJ...580L..11N,Limousin:2007cq}.

\begin{figure}
\begin{center}
\hspace{0.9cm}
\vspace{0.2cm}
\resizebox{6.95cm}{!}{\includegraphics{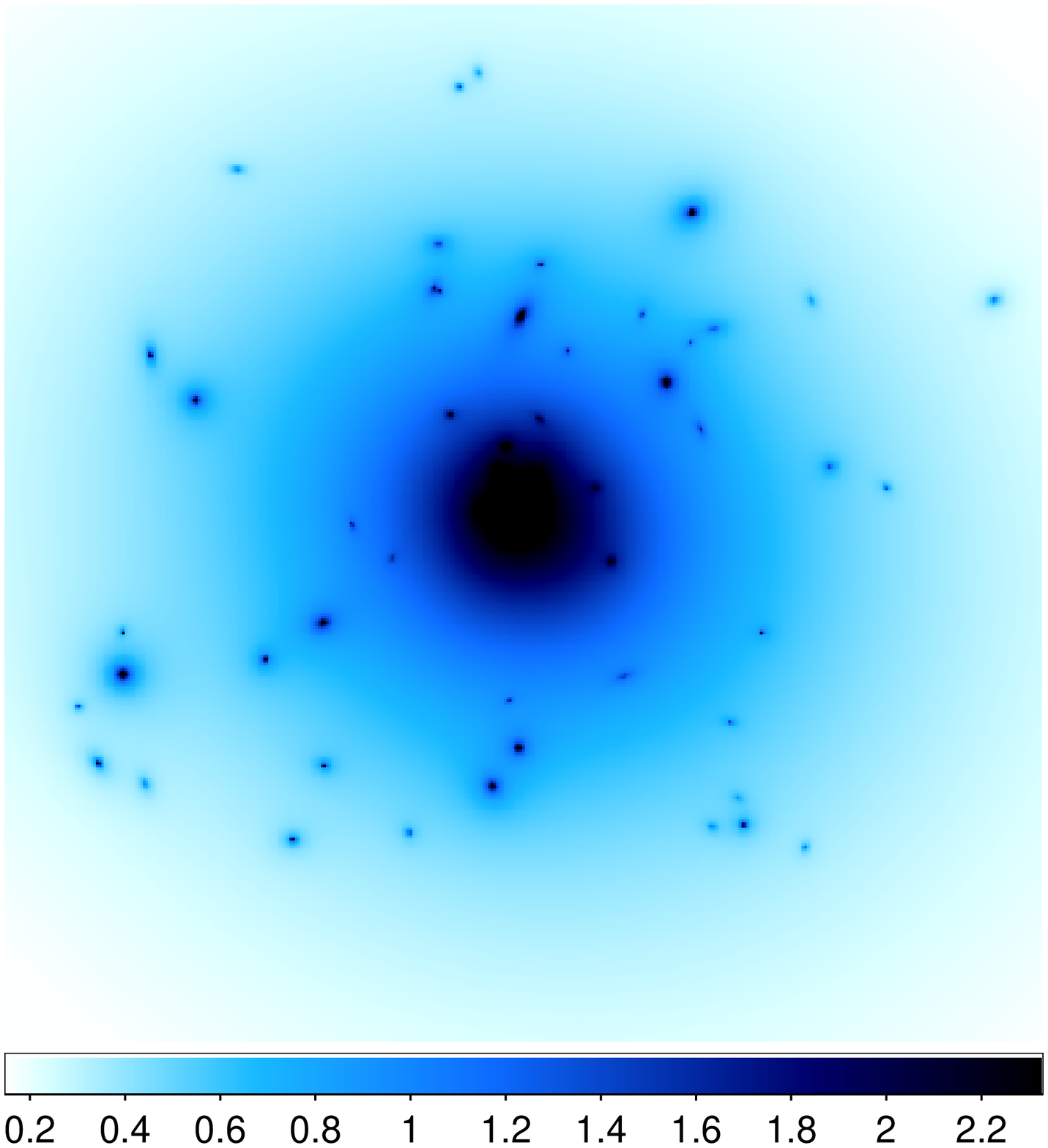}} 
\resizebox{8.4cm}{!}{\includegraphics{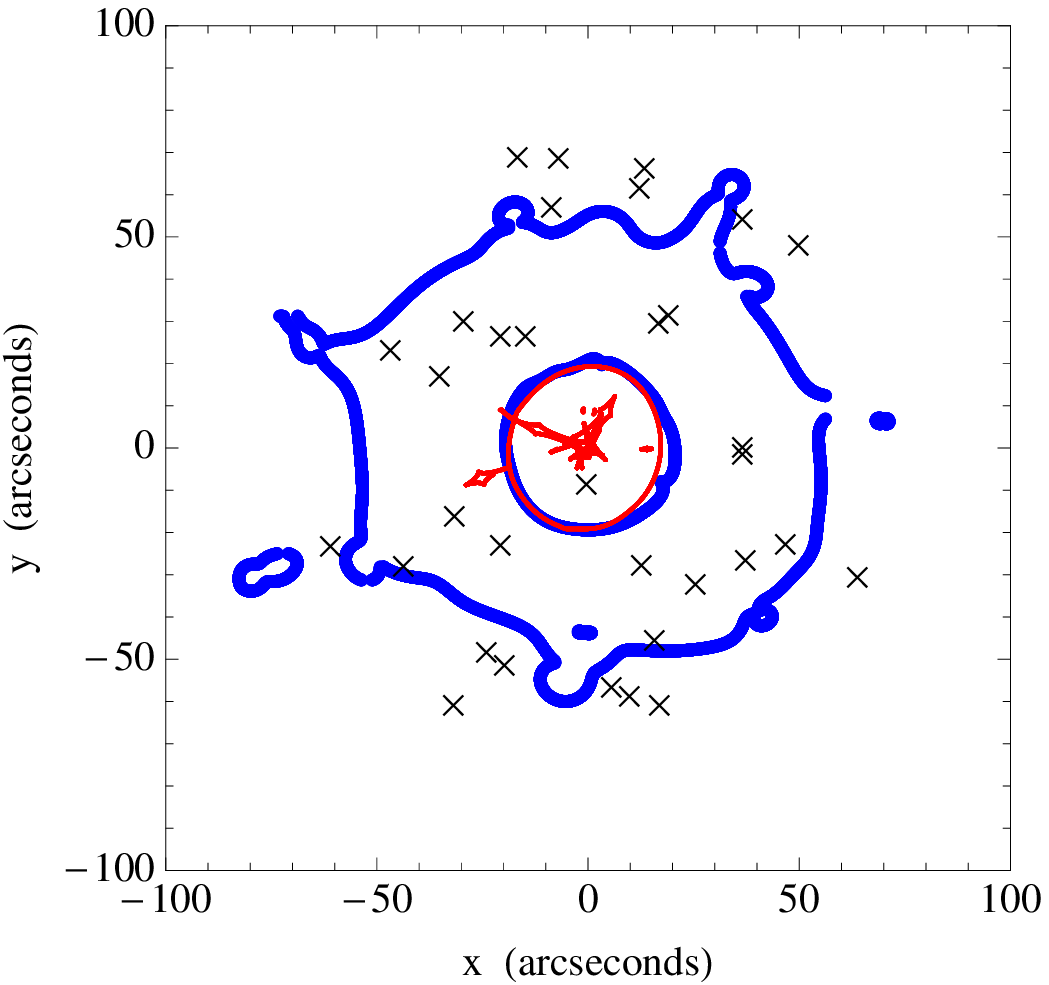}} 
\end{center}
\caption{A simulated lensing configuration with $20$ multiply imaged
families.  Top panel: a convergence map of the lensing cluster with
$z_l = 0.24$ and $z_s = 1.5$.  The size of the frame is $200\times200$
arcseconds$^2$.  Bottom panel: the image positions, shown as x's, and
the critical curves and caustics for $z_s = 1.5$.  The location and
redshifts of images from sources lying between $z_s \sim 0.5-5$ are used to
simultaneously invert the lens and constrain the dark energy
equation-of-state.}
 \label{FIG:examplelens}
\end{figure}  

The first step in simulating a lensing configuration is to create a
smooth, cluster-sized halo.  The parameters $\sigma_v$, $\rcore$, and
$\rcut$ are drawn uniformly from the intervals
$1000-1500~\mathrm{km/s}$, $30-100~\kpc$, and $800-1000~\kpc$
respectively.  An ellipticity and position angle are randomly assigned
in the intervals $0-0.3$ and $0-360$ degrees.  The halo center is
fixed at the origin.  A cluster redshift in the range $z = 0.025-0.6$
is drawn from the distribution in \citet{2009MNRAS.396..354G}, which
was derived from the MAssive Cluster Survey
\citep{2001ApJ...553..668E}[MACS].  In this work, we do not take into
account the effects of dark energy on cluster assembly characteristics.  We
note that a change in cluster properties, such as an increase in their
concentrations as indicated in \citet{2009MNRAS.394.1559G} for
example, could certainly affect the number of multiple-image families
and multiplicities observed.  However, we emphasize that the CSL
technique utilized here is purely geometric in nature, and does not
rely on cluster properties to constrain dark energy.

Galaxy-scale potentials are incorporated into the simulated clusters
with the mean scaling relations derived empirically from observations,

\begin{equation}
\sigma_v = \sigma_v^*\left( \frac{L}{L^*}\right)^{1/4}
\label{sigmascalingEQ}
\end{equation}
\begin{equation}
\rcut=\rcut^* \left( \frac{L}{L^*} \right)^{1/2},
\label{rcutscalingEQ}
\end{equation}
where $L$ is the galaxy luminosity.  Scaling relations such as these
are typically used to model galaxy halos in parametric analyses.  The
scaling of the velocity dispersion with luminosity is motivated by the
Tully-Fisher and Faber-Jackson relations for spiral and elliptical
galaxies respectively.  We fix the core radii of all galaxy sub-halos
to be vanishingly small so that the density profiles are approximately
isothermal ($\rho \sim r^{-2}$) inside of the scale radii.  Although
it is sensitive to the mass of sub-halos, cluster strong lensing is at
present unable to distinguish between different profile shapes for
small-scale structures
\citep{2002ApJ...580L..11N,2009ApJ...693..970N}.

In order to account for cluster-to-cluster variation in the galaxy
scaling relations, for each lens we draw $\sigma_v^*$ and $\rcut^*$
from Gaussian distributions with mean values $\bar{\sigma}_v^* =
200~\mathrm{km/s}$, $\bar{r}_{\mathrm{cut}}^* = 40~\kpc$.  The standard
deviations for the $\sigma_v^*$ and $\rcut^*$ distributions are
$40~\mathrm{km/s}$ and $15~\kpc$ respectively, in accordance with
observational lensing studies \citep{2002ApJ...580L..11N}.  For each galaxy, we
first draw an $I$-band absolute magnitude in the range $M_I = -18$ to
$M_I = -23$ from a Schechter function with $M_I^* = -21.7$ and $\alpha
= -1.3$, consistent with recent cluster observations
\citep[e.g.][]{2009AJ....137.3091H}.  The PIEMD parameters $\sigma_v$
and $\rcut$ are then calculated using (\ref{sigmascalingEQ}) and
(\ref{rcutscalingEQ}).  In section \ref{SEC:modelingerrors}, we will draw individual $\sigma_v$ and $\rcut$ values from Gaussian distributions to account for scatter in the scaling relations (\ref{sigmascalingEQ}) and (\ref{rcutscalingEQ}).  We uniformly draw a position angle and an
ellipticity in the interval $\epsilon = 0 - 0.75$.  Finally, a galaxy
position is drawn from the cluster-scale PIEMD density profile. We
place all galaxies within $300~\kpc$ of the cluster center.  In
addition to 50 galaxies, for each cluster we draw $M_I$ between $M_I =
-23$ and $M_I = -24$ from the above Schechter function for the
Brightest Cluster Galaxy (BCG).  For simplicity, we use the same
scaling relations for the BCG as the other cluster galaxies to assign the
corresponding $\sigma_v$ and $\rcut$.  The BCG position is drawn from a
uniform distribution within a radius of $50 ~\kpc$ from the cluster
center.  The BCG $\sigma_v$ and $\rcut$ parameters are left free
during parameter recovery.

The next step is to randomly draw a unique distribution of background
galaxies to be lensed by each cluster.  We use the redshift
distribution obtained by \citet{2009MNRAS.396..354G} from the WFPC2
Hubble Deep Field and Hubble Deep Field-South photometric redshift
catalogues \citep{1999ApJ...513...34F,2000ApJ...538..493Y}.  We create
a population of background galaxies by drawing redshifts and AB(8140)
apparent magnitudes in the ranges of $z=0.4-5.0$ and $M_{AB}(8140) =
19-28$ respectively from this distribution.  The galaxy positions are
drawn uniformly in the field of view.  The background galaxy number
density is fixed to $215~\mathrm{arcmin}^{-2}$ in accordance with the
Hubble Deep Field as in \citet{2009MNRAS.396..354G}.

After drawing cluster and source properties we lens the population of
background galaxies.  We apply a magnitude cut of $M_{AB}(8140) =
24.5$ to remove simulated images that are in practice too dim to
obtain accurate spectroscopic redshifts.  We then sort the images into
multiple-image families, which consist of at least two detectable
images.  The rest of the images are discarded.  The positions and
redshifts of the final families constitute a fiducial catalogue of
constraints for the lensing system.  In Section \ref{SEC:modelingerrors}, we will explore perturbations to the catalogs due to complexities in the cluster galaxy population and LOS structure. 

Figure~\ref{FIG:examplelens} shows an example of a lensing
configuration generated by the above procedure.  The top panel shows
the convergence map of a cluster at $z_l = 0.24$ with $z_s = 1.5$,
illustrating the smooth cluster component and $51$ core cluster
galaxies.  The field of view is $200$ by $200$ arcseconds$^2$,
consistent with an ACS image.  The bottom panel shows the
corresponding critical curves and caustics for $z_s = 1.5$.  The panel
also shows the image positions for $20$ strongly lensed sources with
redshifts ranging from $z_s = 0.87$ to $z_s = 3.4$.

\subsection{Bayesian MCMC}
\label{SEC:bayesianMCMC}

We use the Bayesian MCMC sampler in the LENSTOOL
software\footnote{http://www.oamp.fr/cosmology/lenstool} for parameter
recovery.  In this section we provide a brief summary of the technique.
More details can be found in \citet{Jullo:2007rt}.  LENSTOOL is employed to map image locations to the source plane and vice versa using equation (\ref{lensEQ}) for a given lens configuration. An image plane $\chi^2$ statistic may be computed from

\begin{equation}
\chi_I^2 = \sum_{f=1}^{N_f}\sum_{i=1}^{n_f}{\left( \vec{\theta}_{fi} - \vec{\theta}^{\mathrm{mod}}_{fi} \right)^T C^{-1}_{fi} \left( \vec{\theta}_{fi} - \vec{\theta}^{\mathrm{mod}}_{fi} \right)}
\label{EQ:implchi2}
\end{equation}
where $\vec{\theta}_{fi}$ and $\vec{\theta}^{\mathrm{mod}}_{fi}$ are
the observed and model positions of image $i$ of family $f$
respectively and $C$ is the corresponding covariance
matrix.  Working in the image plane, particularly in the case where a
large number of $\chi^2$ are computed, is time consuming due to the
fact that the lens equation must be inverted.  A less computationally
expensive approach is to linearize the lens equation and rewrite
equation ($\ref{EQ:implchi2}$) in terms of source plane quantities.  In
this case, 

\begin{equation}
\chi_S^2 = \sum_{f=1}^{N_f}\sum_{i=1}^{n_f}{\left(
\vec{\beta}_{fi} - \left< \vec{\beta}_f \right> \right)^T M^T C^{-1}_{fi} M \left(
\vec{\beta}_{fi} - \left< \vec{\beta}_f \right> \right)},
\label{EQ:soplchi2}
\end{equation}
where $M$ is the magnification tensor, $\vec{\beta}_{fi}$ is the source plane
position, and $\left< \vec{\beta}_f \right>$ is the barycenter of
family $f$.  In the source plane approach, the models with the lowest
$\chi^2$ are the those in which source plane positions from the same
family exhibit minimal scatter.  It is often the case that some images
in a family go undetected due to obstruction or demagnification.  Note
that the location of \emph{all} images in a given family is not
required.  We only require the identification of more than one image
in a family.

The parameters in the models used for the MCMC are summarized in Table
\ref{ParameterTable}.  We use flat 50 per cent priors, centered on the
input values, for all mass profile parameters, except for the cluster
center, which we limit to $\pm10$ arcseconds.  Our results are
insensitive to the width of the cluster center limits.  For
simplicity, we assume that mass traces light for the cluster galaxies,
and fix their positions.  In the case of constant equation-of-state,
we assume flat priors of $0 < \Omega_m < 0.7$ and $-2 < w_x < 0$.  For
the CPL parameterization, we assume flat priors of $-3 < w_0 < 0$ and
$-6 < w_a < -w_0 $.  We also assume $0 < \Omega_m < 0.7$.  In
practice, the exclusion of $w_a > - w_0$ would come from higher
redshift constraints \citep[see][]{Kowalski:2008ek}.  The CPL
parameterization approaches $w_0+w_a$ as $z\rightarrow\infty$.  Hence
models above the $w_a = - w_0$ line have $w_x > 0$ at high redshifts,
implying that dark energy dominates at early times. Current results
favor dark energy making up only a small fraction of the total energy
density for $z \gg 1$ \citep[e.g.][]{Kowalski:2008ek,2010arXiv1001.4538K}.

The MCMC sampler works by randomly drawing parameter sets from Table
\ref{ParameterTable}.  The $\chi^2$ is computed using the simulated
position and redshift data and each model is either accepted or
rejected using a variant of the Metropolis-Hasting algorithm.  As the
process is repeated, the sample of points in the parameter space
converges to the posterior probability density function (PDF).  To
avoid getting stuck in local maxima, the MCMC sampler in LENSTOOL
utilizes the convergence technique of selective annealing.  Following
\citet{Jullo:2007rt}, we use a convergence rate of 0.1 and fix the
number of samples to be $50,000$.

%%%%%%%%%%%%%%%%TABLE 1
\begin{table}
\caption{  Free parameters in the Bayesian MCMC sampler.}
\begin{tabular}{ll}
\hline\hline
Cluster-scale parameters &  Description \\
\hline
\\ [-1ex]
$x$,$y$  & Center of the cluster density profile\\
$\sigma_v$  & Central velocity dispersion \\
$\rcore$ & Core radius \\
$\rcut$ & Scale radius \\
$\epsilon$ & Ellipticity\\
$\theta_{\PA}$ & Position angle\\ [1.5ex]
\hline
BCG\\
\hline
\\ [-1ex]
$\sigma_v$, $\rcut$ & Velocity dispersion and scale radius \\
& of the brightest cluster galaxy \\ [1.5ex]
\hline
Galaxy-scale\\
\hline
\\ [-1ex]
$\sigma_v^*$, $\rcut^*$ & Normalization of the scaling relations \\ & (\ref{sigmascalingEQ}) and (\ref{rcutscalingEQ}) \\ [1.5ex]
\hline
Cosmological \\
\hline
\\ [-1ex]
$\Omega_m$ & Mean matter density \\
$w_x$ & Dark Energy equation-of-state  \\
or & \\
$w_0$,$w_a$ & \\[1.5ex]
\hline
\\
\end{tabular}

\label{ParameterTable}
\end{table}

%%%%%%%%%%%%%%%%%%%%%%%%%%%

\section{Modeling errors in CSL}
\label{SEC:modelingerrors}

\subsection{Effect of cluster galaxies}
\label{SEC:cgscatter}

The inclusion of galaxy-scale potentials in cluster lens models
greatly improves their ability to reproduce observed images.  In some
cases, individual cluster galaxies can have a large impact on the
configuration of images, and it may difficult to reproduce
observations without modeling them individually (for example, when the
presence of a cluster galaxy changes the multiplicity of images).
However, regarding the bulk of the galaxy population, one is forced to
make some simplifying assumptions owing to the limited number of
constraints and large number of parameters required.  Specifically,
the velocity dispersion and scale radii parameters are typically
assumed to follow empirically motivated scaling relations such as
(\ref{sigmascalingEQ}) and (\ref{rcutscalingEQ}).  In this section we
examine the effect that these simplifying assumptions have on our
ability to reproduce image locations.

\begin{figure}
\begin{center}
\resizebox{8.0cm}{!}{\includegraphics{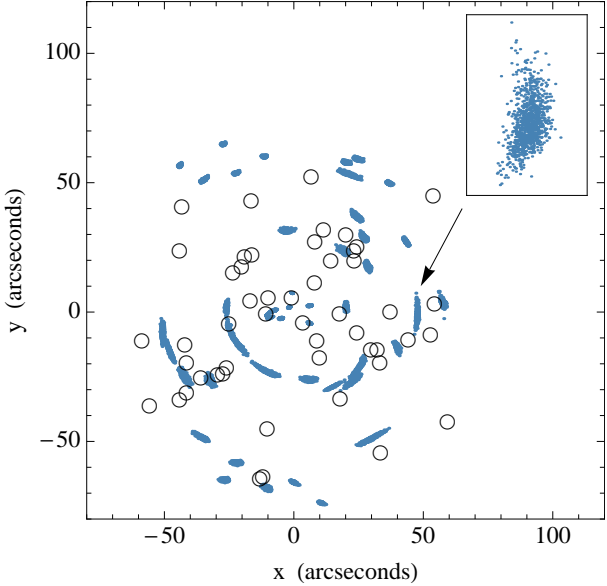}}
 \end{center}
\caption{Results from a Monte-Carlo simulation of CSL illustrating the
effect of scatter in cluster galaxy scaling relations.  We show the
lensed image locations obtained from $1000$ random realizations of the cluster galaxy
population for $z_l = 0.34$.  A $20$ per cent scatter in
the scaling of velocity dispersion and scale radius with luminosity is
assumed.  The open circles show the location of cluster galaxies.  The inset ($6\times16$ arcseconds$^2$) shows the Monte-Carlo realizations for a single image.  Each point corresponds to the location of the lensed image for a particular realization of the cluster galaxy population.  }
\label{FIG:CGscatter}
\end{figure}

It is well known that the Faber-Jackson and Tully -Fisher relations
exhibit considerable scatter.  We should therefore expect significant
scatter in the scaling of $\sigma_v$ with luminosity.  It is also
reasonable to expect a similar degree of scatter in the $\rcut$
relation.  The deviations of individual cluster galaxies from the
relations can introduce perturbations to nearby images which cannot be
accounted for with a simple parametric model.  We should not expect
our models to reproduce images to within this perturbation scale.  We
perform simple Monte-Carlo simulations to quantify these deviations.

\begin{figure}
\begin{center}
 \resizebox{8.5cm}{!}{\includegraphics[angle=-90]{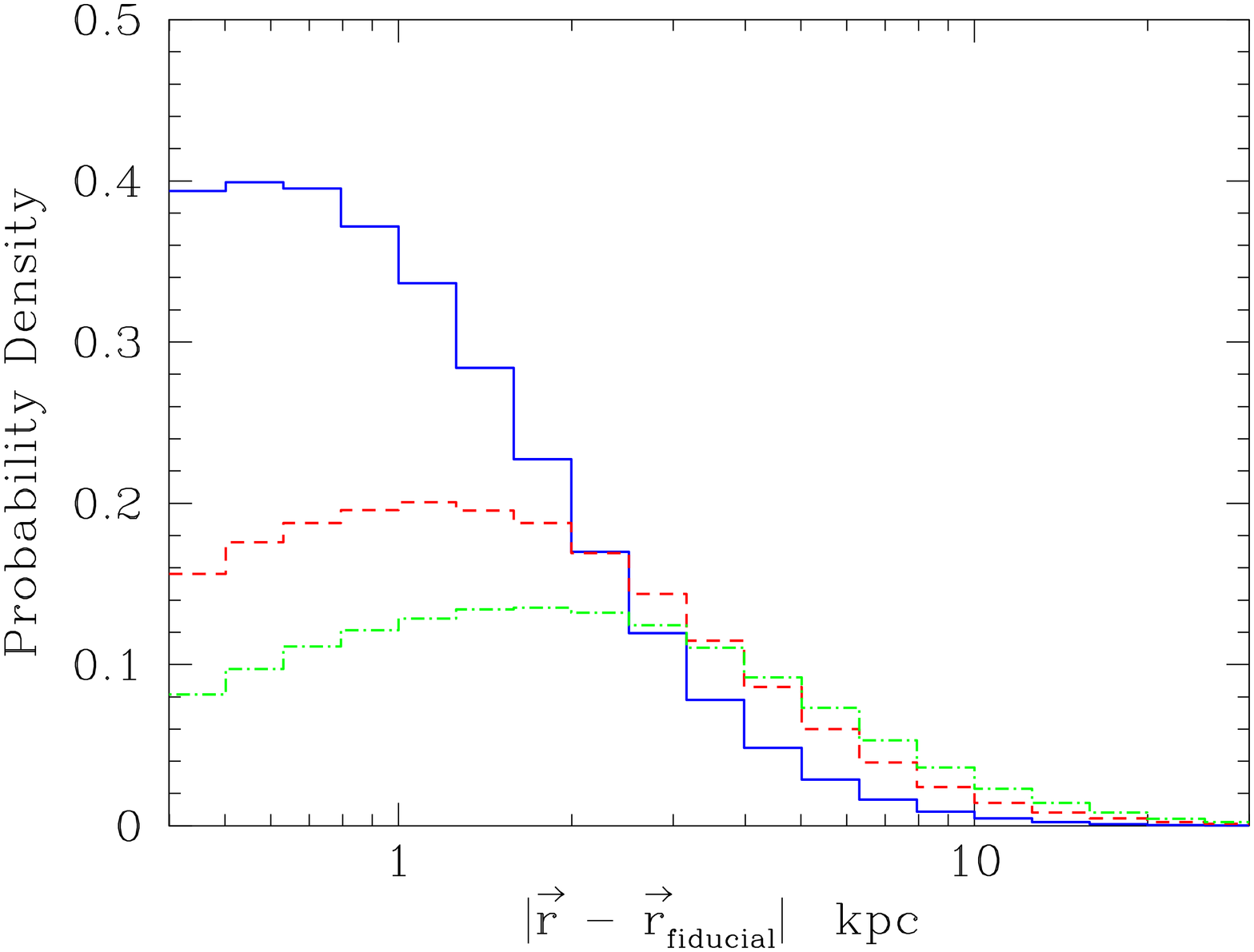}}
 \resizebox{8.5cm}{!}{\includegraphics[angle=-90]{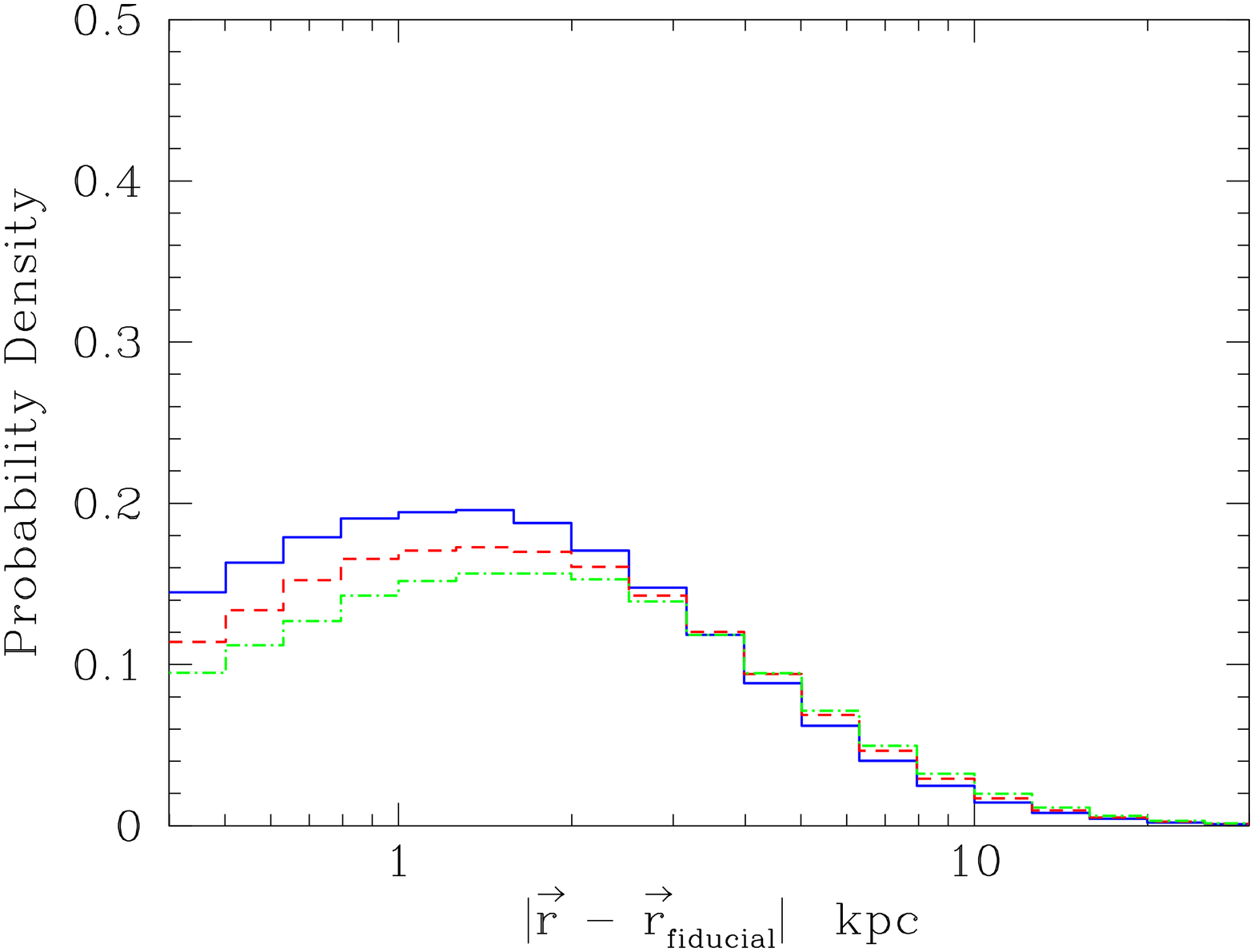}}
 \end{center}
\caption{Top panel: magnitude of image deviations, in comoving $\kpc$,
between models with and without scatter in the cluster galaxy scaling
relations.  For the models with scatter, velocity dispersions and
scale radii are drawn from Gaussian distributions with mean given by
equations (\ref{sigmascalingEQ}) and (\ref{rcutscalingEQ}).  To
illustrate the effect of varying degrees of scatter, we assume
standard deviations of $10$, $20$, and $30$ per cent of the mean
values for the solid, dashed, and dot-dashed histograms respectively.
The scale of these perturbations is used to quantify modeling errors
that arise from complexities in the cluster galaxy population.  Bottom
panel: deviations due to variation in the slope of the scaling
relations, with a fixed $20$ per cent scatter in the $\sigma_v$ and
$\rcut$ values. Assuming mean values of $1/4$ and $1/2$ respectively,
power-law indices for equations (\ref{sigmascalingEQ}) and
(\ref{rcutscalingEQ}) are randomly drawn.  The indices have a minimal
effect due to the scatter in $\sigma_v$ and $\rcut$.}
\label{FIG:CG}
\end{figure}

For each of our simulated lensing configurations, we draw $1000$
different galaxy populations.  We randomly draw velocity dispersions
$\sigma_v$ and scale radii $\rcut$ from Gaussian distributions with
mean values obtained through (\ref{sigmascalingEQ}) and
(\ref{rcutscalingEQ}) in the fiducial models.  We use standard
deviations of $10$, $20$, and $30$ per cent of the mean to explore how
varying degrees of scatter affect image positions.  We lens the given
source catalog through each realization and examine the scatter in
image positions.  Figure~\ref{FIG:CGscatter} shows results from one
such lensing configuration.  Each point corresponds to an image in one
realization of the Monte-Carlo simulations.  The open circles indicate the
position of cluster galaxies.  The tangential shapes of the image
distributions illustrate that the deviations are generally larger
along the local principal magnification direction.

In the top panel of Figure~\ref{FIG:CG}, we compute image deviations
relative to the fiducial models with no scatter in the cluster
galaxies, and show the probability density of these deviations.  The
solid, dashed and dot-dashed histograms correspond to $10$, $20$, and $30$
per cent scatter respectively.  Since each simulated cluster in our
sample is at a different redshift, we calculate the deviations in
comoving $\kpc$. The average deviation in each case is $2.5$, $4.9$
and $6.9$ comoving $\kpc$ respectively.  This corresponds to $0.6$,
$1.3$, and $1.8$ arcseconds for clusters at $z \sim 0.2$ respectively.
In rare cases, deviations can be as large as $\sim10$ arcseconds.

We now turn our attention to the effect of the power law indices in
equation (\ref{sigmascalingEQ}) and (\ref{rcutscalingEQ}).  Although
these relations have been successfully used in the past for
constraining the mass distributions of cluster lenses, we examine here
more closely the implications of these assumptions. As discussed
above, the scaling of $\sigma_v$ with $\sim L^{1/4}$ is motivated by
well established empirical results. However, while $\sigma_v$ itself
is not directly measured observationally, studies of individual galaxy
lenses from the SLACS survey for instance \citep[see][]{2008ApJ...684..248B,2009MNRAS.399...21B} suggest that the
measured stellar velocity dispersion within an effective radius is a
good proxy for the velocity dispersion of the lensing mass
model. Owing to a lack of empirical data, the assumed scaling of $\rcut$ with luminosity is more
uncertain and may vary considerably.

To determine the effect of assuming an incorrect scaling with
luminosity, we again perform Monte-Carlo simulations.  For each
cluster in our sample, we draw 1000 realizations where we vary the
power-law indices.  We draw from a Gaussian with mean given by $1/4$
($\sigma_v$) and $1/2$ ($\rcut$).  We again use standard deviations of
10, 20, and 30 per cent of the mean to explore the effect of various
amounts of scatter.  We assume a fixed scatter in the $\sigma_v$ and
$\rcut$ values themselves of 20 per cent.  The bottom panel of Figure~\ref{FIG:CG} 
shows image deviations with respect to the fiducial
models in which there is no scatter in the scalings whatsoever.  The
10, 20, and 30 per cent cases are shown.  We note that varying the
power law indices has only a mild effect on the deviations if there is
scatter present in the values themselves.  The effect of the indices
is essentially lost in the scatter.

Finally, we have checked that the above deviations are relatively
independent of the cosmology, varying only by a few per cent for a
wide range of $\Omega_m$ and $w_x$ parameters.

 \subsection{The effect of line-of-sight halos}

\subsubsection{The multiple lens plane approximation}

The thin-lens approximation, equation (\ref{lensEQ}), applies to the case where the lensing mass distribution is localized.  In this case, the deflection angle may be approximated as a single impulse.  However, in cosmological applications, light from distant sources may be deflected by many structures along the path to the observer.  The matter distribution may be broken up into multiple lens planes so that the deflection angle is approximated by a series of impulses.  In this case, the angular position of the light ray at lens plane $j$ is given by  

\begin{equation}
\vec{\theta}_j = \vec{\theta}_1 - \sum_{i=1}^{j-1}{\frac{D_{ij}}{D_j}~\hat{\alpha}\left(\vec{\theta}_i\right)},
\label{multlensEQ}
\end{equation}
where $\vec{\theta}_1$ is the observed image position and $\hat{\alpha}\left(\vec{\theta}_i\right)$ is the deflection angle evaluated at the ray position on plane $i$ \citep[see][for example]{1992grle.book.....S}.  Note that the lens plane iterator increases with distance from the observer.  Equation (\ref{multlensEQ}) may be iterated to trace an image location back to the appropriate source plane. 

Following \cite{Hilbert:2008to}, we use a quicker and more computationally efficient alternative to equation (\ref{multlensEQ}), where the ray position on plane $j$ is written in terms of the positions on lens planes $j-1$ and $j-2$,

\begin{eqnarray}
\vec{\theta}_{j} & = & \left( 1- \frac{f_K^{(j-1)}}{f_K^{(j)}} \frac{f_K^{(j-2,j)}}{f_K^{(j-2,j-1)}} \right)\vec{\theta}_{j-2} \nonumber \\ & & +\frac{f_K^{(j-1)}}{f_K^{(j)}}\frac{f_K^{(j-2,j)}}{f_K^{(j-2,j-1)}}\vec{\theta}_{j-1}  \label{multilensEQB}
\\ & & -\frac{f_K^{(j-1,j)}}{f_K^{(j)}}\hat{\alpha}\left( \vec{\theta}_{j-1}\right). \nonumber
\end{eqnarray}
Here, $f_K$ is the comoving angular diameter distance, $f_K^{(a,b)} \equiv D(z_a,z_b)/(1+z_b)$.

We developed our own software to perform the ray tracing calculations in the following sections.  At present, the code can perform ray tracing through planes consisting of lenses with analytic mass profiles, such as the PIEMD and NFW profiles.  For the task of obtaining image locations, the code traces grid points backwards to the appropriate source redshift using equation (\ref{multilensEQB}).  The set of transformed grid points in the source plane is then divided into triangles \citep[c.f.][]{1992grle.book.....S}.  The triangles enclosing the source are identified and refined by repeatedly shrinking their sizes to improve the resolution of the image location.

\subsubsection{Correlated halos}
\label{SEC:correlatedhalos}

In this section, we explore the impact of structures situated along
the line-of-sight in cluster strong lensing systems.  We begin by
considering, in a simplified manner, the effects of correlated
structures.  In the next section, we will consider uncorrelated halos
in the light cone.

All CSL studies to date have utilized a single lens plane for mass
modeling.  In what follows, we ask whether correlated structures can
be reasonably modeled on the same redshift plane as the cluster, or
whether models consisting of multiple lens planes are necessary.  We
consider the simple example of a cluster lens with an infalling group-sized halo residing slightly behind it.  Due to its proximity,
the projection might result in the association of the group-sized halo
with the cluster system.  We simulate a cluster at $z = 0.192$ with an aperture mass of
$M = 9.5 \times 10^{14}~\Msun$ within a radius of 1 Mpc.  The cluster has $40$ core galaxies containing a total mass of $2 \times 10^{13}~\Msun$.  The infalling group, with an aperture mass of $6.3 \times 10^{13}~\Msun$, is located $20$ comoving $\Mpc$
behind the cluster at a redshift of $z=0.198$, and is situated at
$\left(x,y\right) = \left(40,40\right) $ arcseconds with respect to
the cluster center. We associate 10 galaxies with the group.

We randomly draw $20$ different source catalogs with source redshifts
between $z \sim 0.7 - 5$ and lens them with the two-plane model.
Multiply imaged systems are identified and selected.  To test the
effect of using only one lensplane to model the system, we create a
model in which the group-sized halo and associated galaxies reside on
the same lensplane as the cluster.  We lens the same $20$ catalogs and
compare image locations.

\begin{figure}
\begin{center}
\resizebox{8.0cm}{!}{\includegraphics{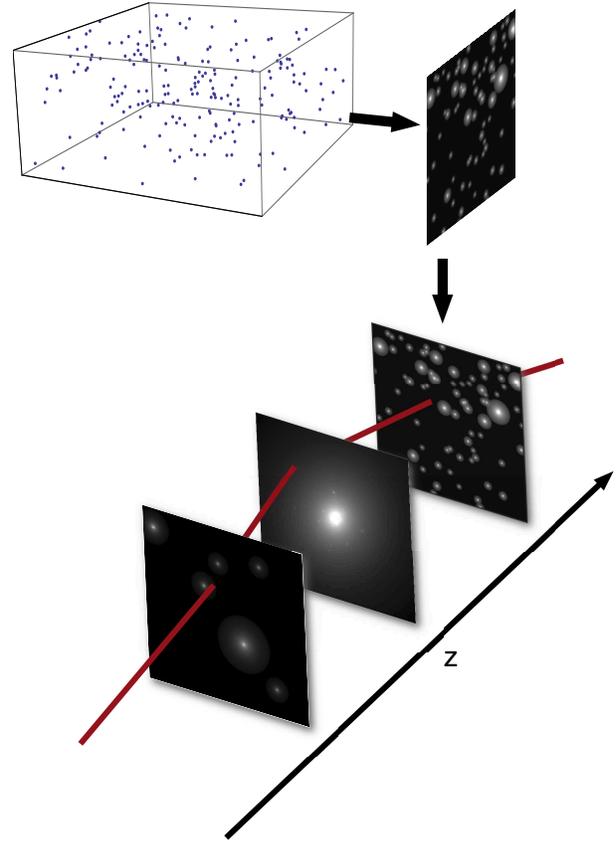}}
 \end{center}
\caption{Schematic diagram illustrating the creation of lensplanes to
quantify the effects of LOS halos. A rectangular slice of the
Millenium Simulation box is taken. The locations of halos are
projected along the long axis and analytic NFW potentials are placed
on those sites.  The NFW parameters are obtained through scaling
relations with mass and redshift.  The lensplane is inserted at the
appropriate redshift and a multi-plane lensing algorithm is used to
trace rays. In this work, forty-two lensplanes are used between $z=0$
and $z=5$.}
\label{FIG:LOScartoon}
\end{figure}

\begin{figure}
\begin{center}
\resizebox{8.0cm}{!}{\includegraphics{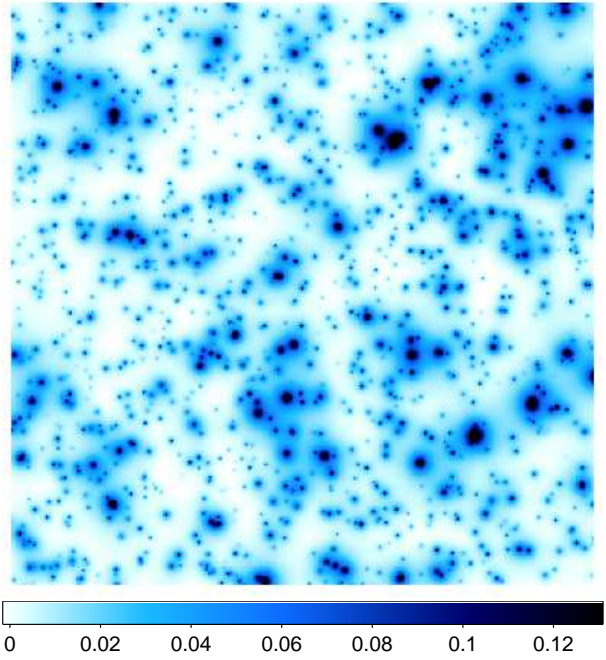}}
 \end{center}
\caption{Example convergence map of a simulated line-of-sight created
from Millenium Simulation halo catalogs, for $z_s = 1.5$.  The field
of view is $400 \times 400$ arcseconds$^2$.  For each simulation
snapshot, a lensplane is created by projecting halo locations and
placing analytic NFW potentials.  The NFW parameters are obtained
through scaling with the mass and redshift of the halo.  We use many
line-of-sight realizations such as the above to calculate
perturbations from the cluster-only model, and quantify modeling
errors. }
\label{FIG:convergence}
\end{figure}
 
We find the mean deviation between the two catalogs to be $0.26$
arcseconds, with largest deviations of $\sim 1$ arcsecond.  Such cases
are rare - $87(97)$ per cent of deviations are less than $0.5(1)$
arcseconds - and typically occur for images located close to the
group-sized halo.  We also tested a single lensplane model which is
located halfway between the cluster and group-sized halo.  We found
almost no difference in the deviations.  Based on the magnitude of
deviations, we conclude that in the case where there is a dominant
lensing system in mass, one can reasonably use a single lensplane
model.  Such an approach may introduce modeling errors which are
typically on the order of a few tenths of an arcsecond.  However, we
note that these effects are subdominant to the effects of scatter in
the cluster galaxy population, as shown in the last section.
Moreover, in the next section, we will show that the effects of
uncorrelated galaxy-scale halos along the line-of-sight are greater.
 
We now consider the case when the masses of the two line-of-sight
structures are roughly equivalent.  We set the mass of the two halos
to $4.5 \times 10^{14}\Msun$ and lens the $20$ source catalogs again.
The deviations increase to $0.91$ arcseconds on average and can be as
high as a few arcseconds.  We conclude that in the case with two
equal-sized halos, it may be necessary to utilize two lensplanes in
order to minimize modeling errors.
  
We note that the above investigation is limited due to the simplified
parametric models used.  Galaxy clusters are typically part of a rich
network of structures which cannot be fully captured in the approach
we have taken.  To test these effects in a more realistic way,
including larger-scale structures such as filaments, it is best to
utilize particle data from a cosmological N-body simulation.  However,
such an investigation is beyond the scope of the current work, and we
defer to a future paper.

\subsubsection{Uncorrelated halos}
\label{SEC:uncorrelatedhalos}

\begin{figure}
\begin{center}
\resizebox{8.5cm}{!}{\includegraphics[angle=-90]{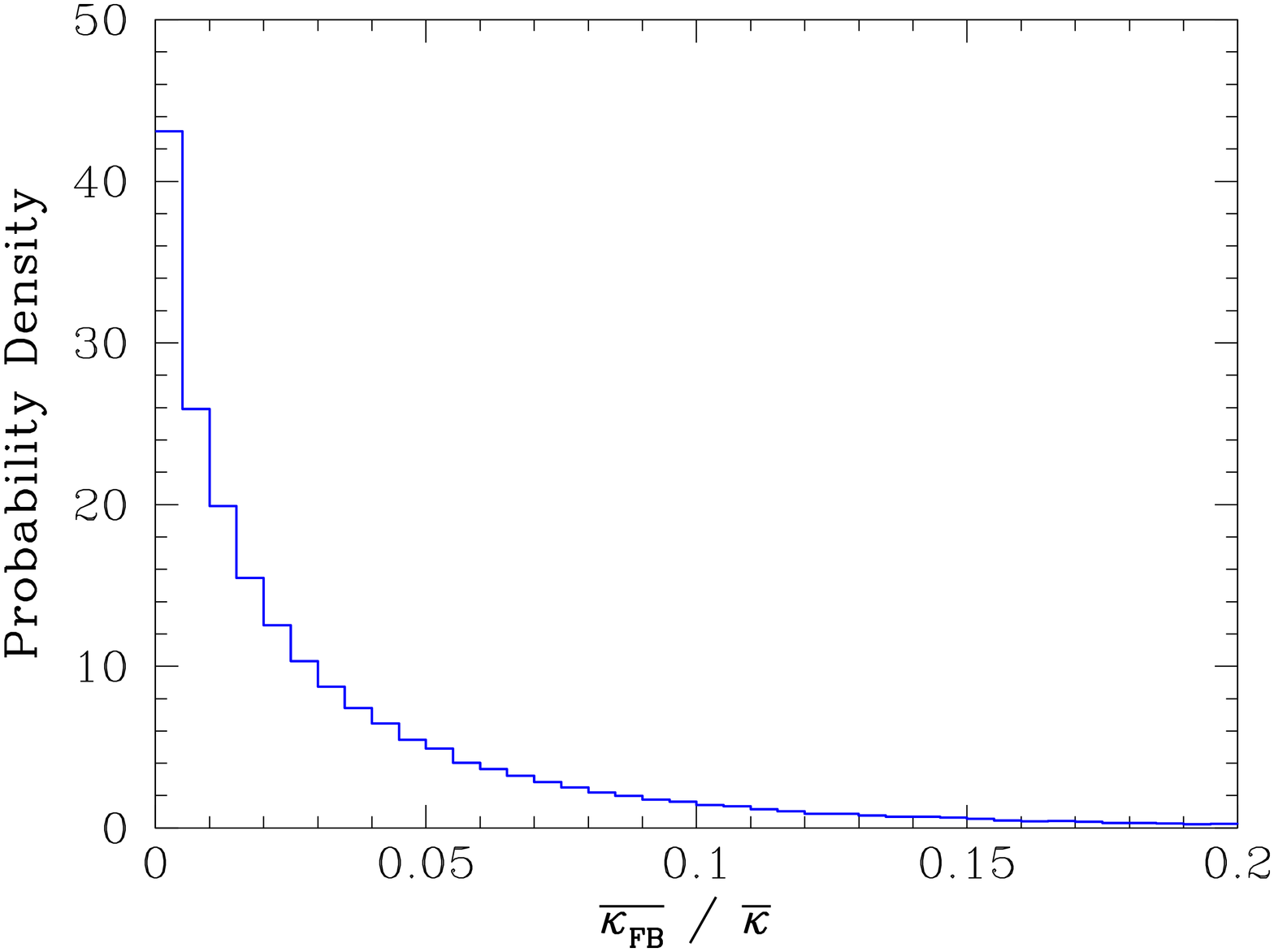}}
 \resizebox{8.5cm}{!}{\includegraphics[angle=-90]{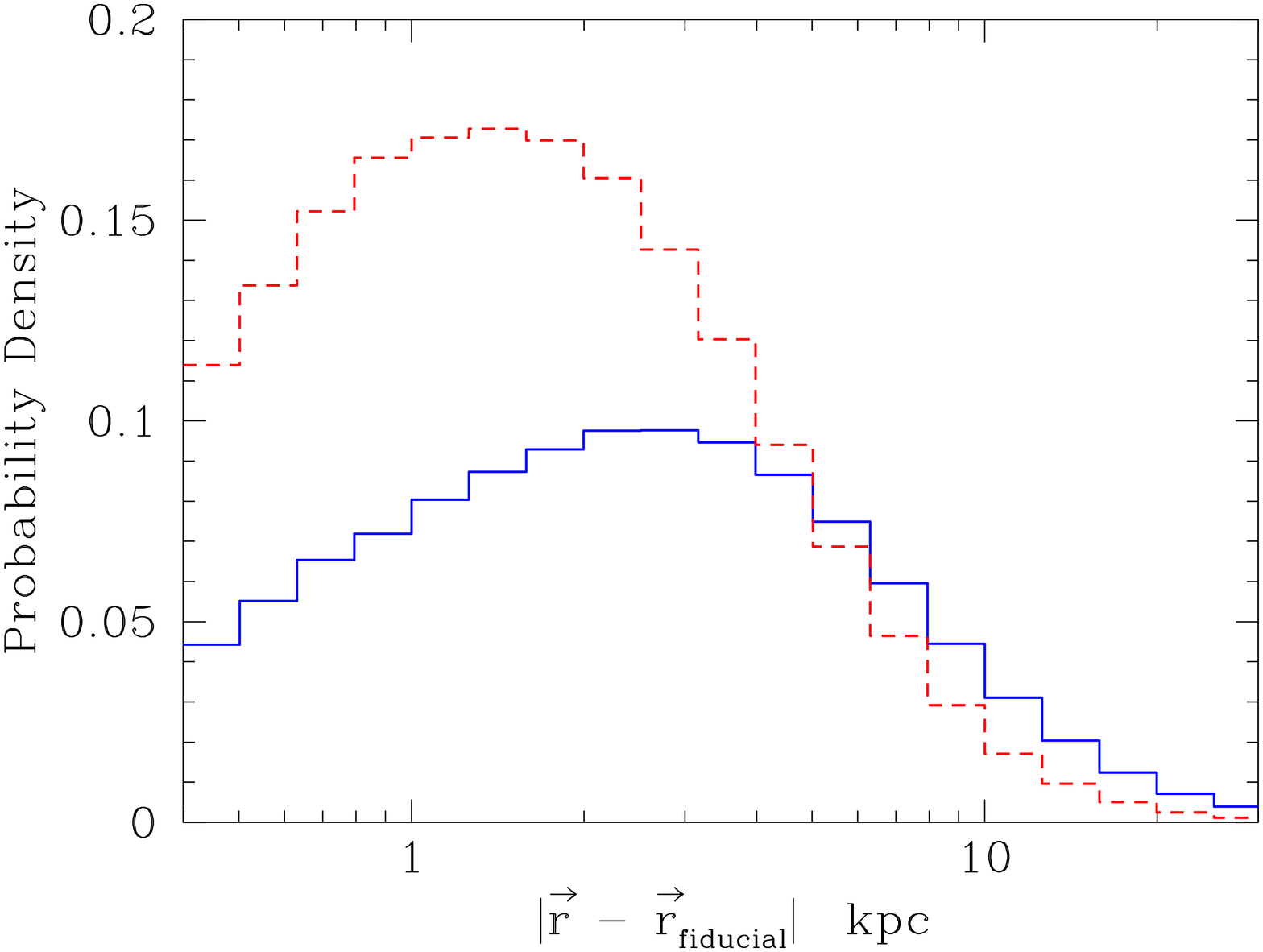}}
 \end{center}
\caption{Top panel: ratio of LOS convergences excluding ($\bar{\kappa}_{\mathrm{FB}}$) and including ($\bar{\kappa}$) the contribution from the cluster plane. The contribution from foreground and background halos to the total LOS convergence is typically less than $\sim10$ per cent.  Bottom panel: the solid histogram shows image deviations, in comoving $\kpc$, between the full light-cone and
cluster only models.  For reference, the dashed histrogram
shows the cluster galaxy result for $20$ per cent scatter in the
scaling relations (see Figure~\ref{FIG:CG}).  Uncorrelated LOS halos perturb image locations by $\sim 9$ $\kpc$ on average, corresponding to $2.3$, $1.6$, and $1.2$ arcseconds at $z = 0.2$, $0.3$ and $0.4$ respectively.}
\label{FIG:uncorrelatedLOS}
\end{figure}

We now turn our attention to the influence of uncorrelated
galaxy-scale halos along the line-of-sight. Although direct modeling of LOS
halos from observations may ultimately improve the fit to images,
perhaps a more practical approach at the present is to quantify the
errors due to neglecting them in the lens model.  We utilize halo
catalogs from the Millennium Simulation
Database\footnote{http://www.mpa-garching.mpg.de/millennium/}
\citep{2005Natur.435..629S} and the multiple-lensplane approximation
\citep[see][for example]{1992grle.book.....S}. A schematic of the
procedure is shown in Figure~5.

We create one lensplane for each snapshot between $z = 0$ and $z = 5$,
yielding a total of 42 planes.  For each lensplane, we take a randomly
oriented 3-dimensional slice of the corresponding snapshot (see Figure~\ref{FIG:LOScartoon} for an illustration).  The slices are
rectangular with length equal to $\{r_{\mathrm{com}}(z_{i-1},z_i)+r_{\mathrm{com}}(z_{i},z_{i+1})\}/2$, where $z_i$ is the redshift of snapshot $i$ and $r_{\mathrm{com}}(z_a,z_b)$ is the comoving distance between $z_a$ and $z_b$.  The widths are chosen so that each plane fills a $400
\times 400$ arcsecond$^2$ area after projection. The locations of all
halos above a threshold of $10^{11}~\Msun$ in the slices are projected
along the long axis. The masses and redshifts in the catalog are used
to obtain mean concentration parameters through scaling relations
obtained by \citet{2008MNRAS.387..536G}.  In order to account for
scatter about these relations, we draw concentrations from a
log-normal distribution with standard deviation given by 0.14 \citep{2007MNRAS.381.1450N}.
Circularly symmetric NFW potentials with the corresponding parameters
are then placed at the location of each halo.  Since the mass of the
NFW profile is not convergent, we truncate each profile at the virial
radius \citep[for lensing properties of the truncated NFW profile, see
][]{DAloisio:2008kx}.  A typical convergence map for a simulated LOS,
without cluster, is shown in Figure~\ref{FIG:convergence}.

Since the slices are much smaller than the Millennium Simulation box, we can perform the
above procedure many times for each lensplane to create an ensemble.
We create random LOS realizations in order to quantify the modeling
errors through a Monte-Carlo approach.  We find that $500$ LOS realizations are sufficient for numerical convergence.  For each lensing configuration
in our cluster sample, we lens the source catalogs through $500$ LOS
realizations with the cluster model placed at the appropriate
redshift.  We utilize a ray tracing code that we developed for this
purpose.  

Following \citet{2007MNRAS.382..121H}, we calculate the ``LOS convergence" $\bar{\kappa}$ in order to quantify the lensing effect of the LOS halos.  Note that this is not equivalent to the multi-plane generalization of the convergence, but is nonetheless useful for quantifying the effects of LOS structure.   We ray trace each image in our simulations backwards to their respective source planes.  We sum the surface mass densities weighted by the appropriate lensing efficiencies at each point along the trajectory.  We do not subtract off the mean density of each plane.  For each lensing system, we do this for all 500 LOS realizations.  We calculate the LOS convergence including ($\bar{\kappa}$) and excluding ($\bar{\kappa}_{\mathrm{FB}}$) the contribution from the cluster plane.  The top panel of Figure~\ref{FIG:uncorrelatedLOS} shows the distribution of $\bar{\kappa}_{\mathrm{FB}}/\bar{\kappa}$ values.  The contribution of foreground and background halos to the LOS convergence is typically less than $\sim10$ per cent.    

The LOS halos perturb the image locations in a manner similar to the results shown in Figure~\ref{FIG:CGscatter}, though to a larger extent.  We note that,
in some cases, LOS halos can change the multiplicity of images.  We compare the results of the LOS Monte-Carlo simulations to the image catalogs in the cluster-only cases.  The solid histogram in the bottom panel of Figure~\ref{FIG:uncorrelatedLOS} corresponds to deviations
between the full LOS and cluster-only cases.  For reference, the
dashed histogram shows the case of $20$ per cent scatter in the
cluster galaxy $\sigma_v$ and $\rcut$ parameters (see the top panel of
figure \ref{FIG:CG}).  Figure~\ref{FIG:uncorrelatedLOS} indicates that
the uncorrelated halos in the LOS perturb image locations by $\sim9$
comoving kpc on average (corresponding to $2.3$, $1.6$, and $1.2$
arcseconds at z = $0.2$, $0.3$, and $0.4$ respectively).

\section{Simulated CSL constraints on dark energy}  
\label{SEC:simulatedconstraints}

\begin{figure}
\begin{center}
\resizebox{8.0cm}{!}{\includegraphics{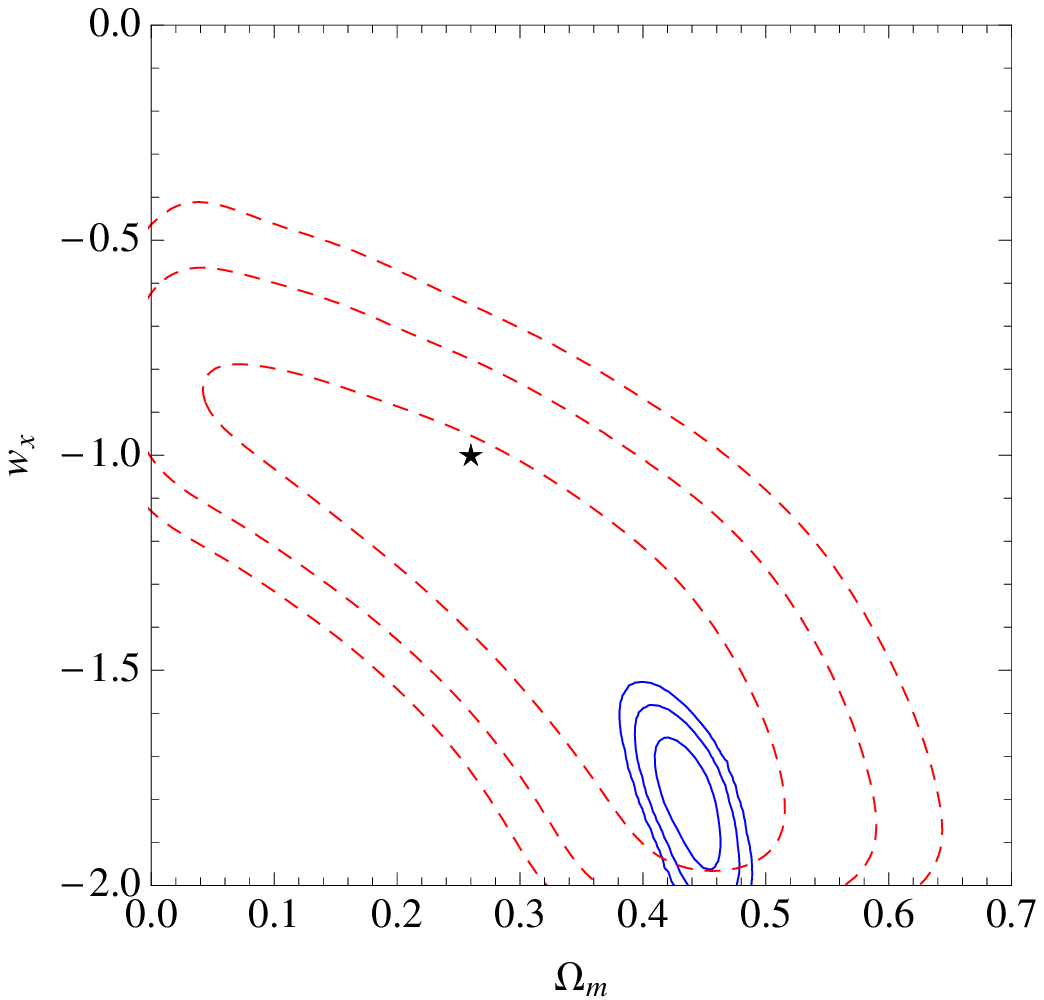}}
\resizebox{8.0cm}{!}{\includegraphics{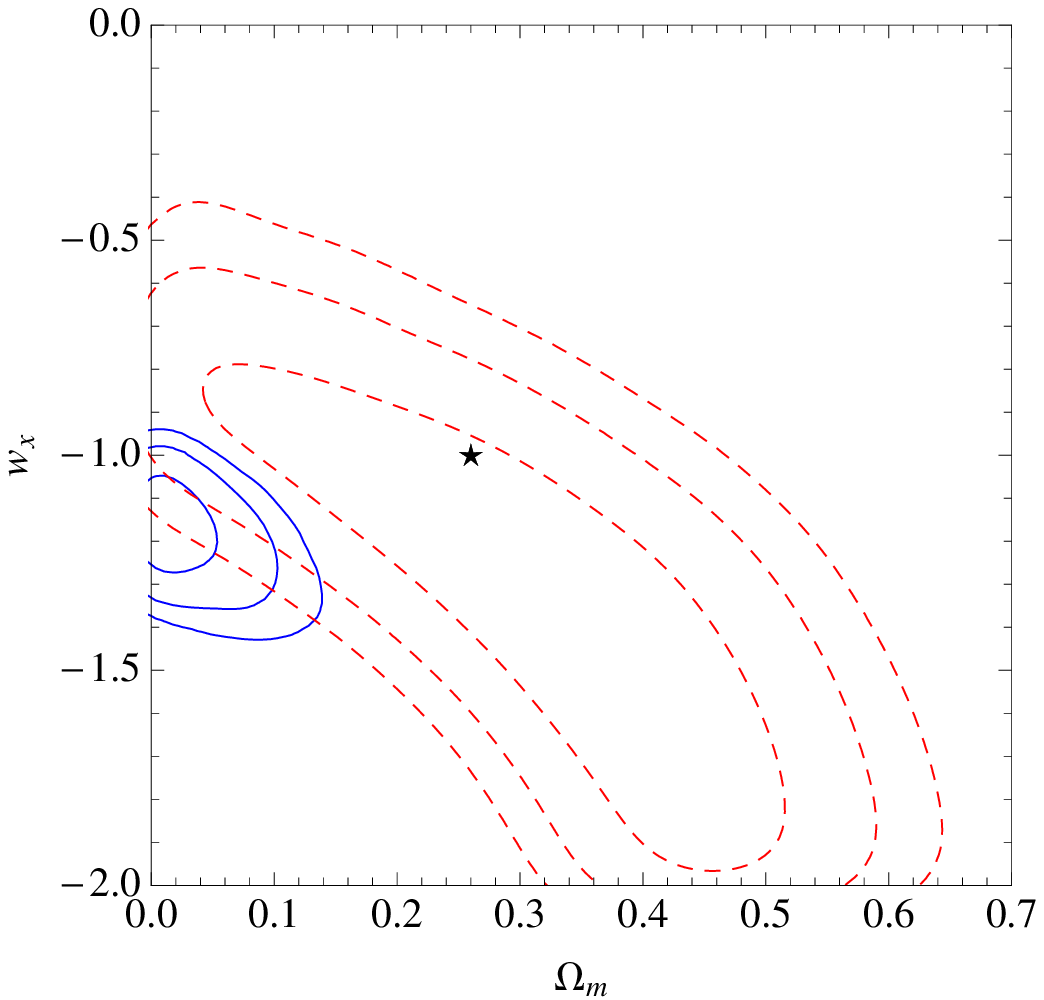}}
 \end{center}
\caption{The effect of underestimating errors in CSL for a single
cluster.  Top panel: the solid contours
assume astrometric uncertainties of only $0.1$ arcseconds on the
image positions, neglecting all sources of modeling error.  The dashed contours correspond to parameter recovery using
the full error estimate, accounting for perturbations due to scatter
in the cluster galaxy population, LOS halos, and astrometric errors. The input model is depicted by a star.  Underestimating the total
error can lead to erroneous constraints on the dark energy
equation-of-state.  Bottom panel: the solid contours show the effect of using photometric redshifts for the high-redshift half of the image catalog.  For these
images, photometric redshift errors are simulated by drawing redshifts from a Gaussian distribution with standard deviation of
$0.5$.  The full error estimates are used for all images.  The dashed contours are the same as in the top panel. }
\label{ImageRedshiftPL}
\end{figure}

Having explored some of the most significant sources of error in CSL,
we now use our simulations to perform a feasibility test on whether
CSL systems with perturbations due to the LOS and cluster galaxy
populations can yield useful dark energy constraints.  We use the set
of ten simulated clusters with $20$ families each discussed above.
Each cluster is placed within an independent light cone generated from
the Millennium Simulation halo catalogs (see Section
\ref{SEC:uncorrelatedhalos}).  The 50 core galaxies for each
cluster are simulated with a $20$ per cent scatter in the luminosity
scaling relations (see Section \ref{SEC:cgscatter}).  Since we have
shown that variation in the power-law indices of equations
(\ref{sigmascalingEQ}) and (\ref{rcutscalingEQ}) makes little
difference in the image perturbations when there is significant
scatter in the galaxy population, we fix them to the input values.

\begin{figure}
\begin{center}
\resizebox{8.0cm}{!}{\includegraphics{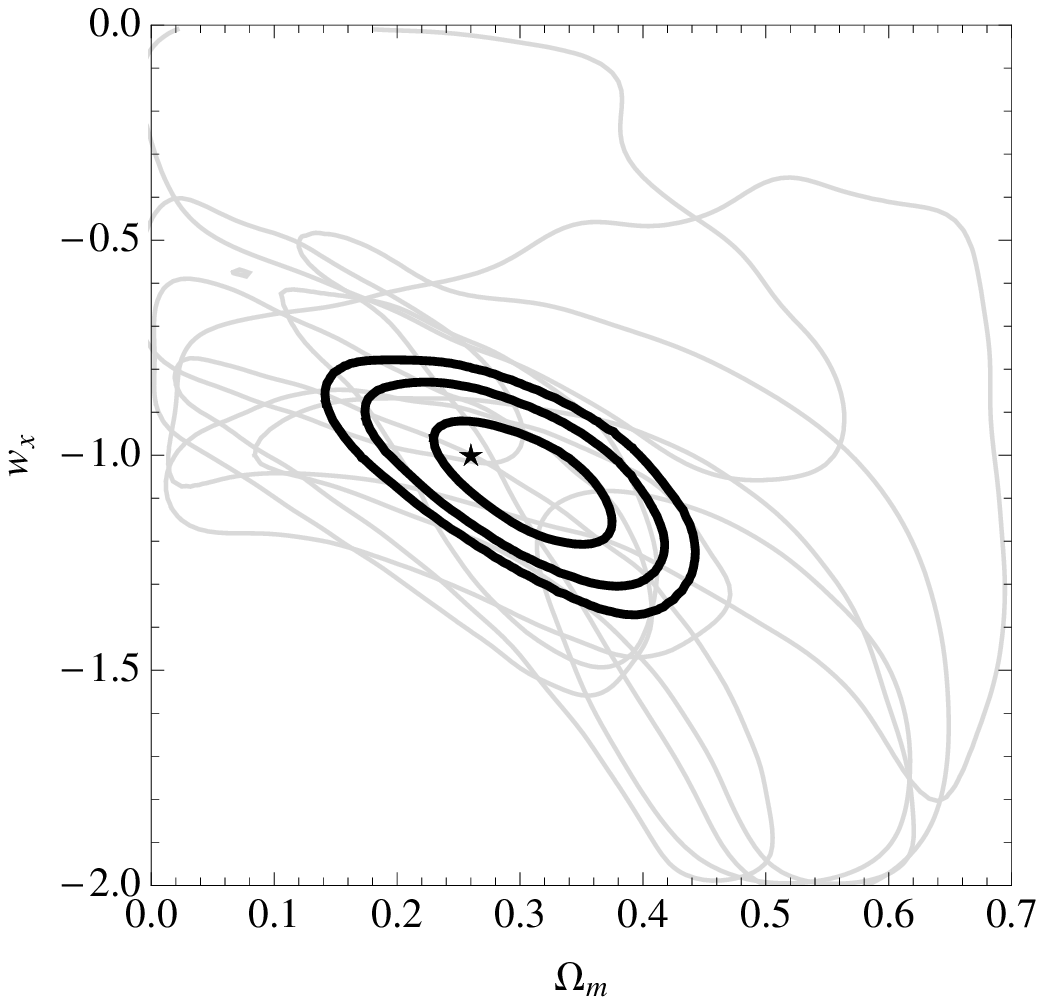}} 
\resizebox{7.8cm}{!}{\includegraphics{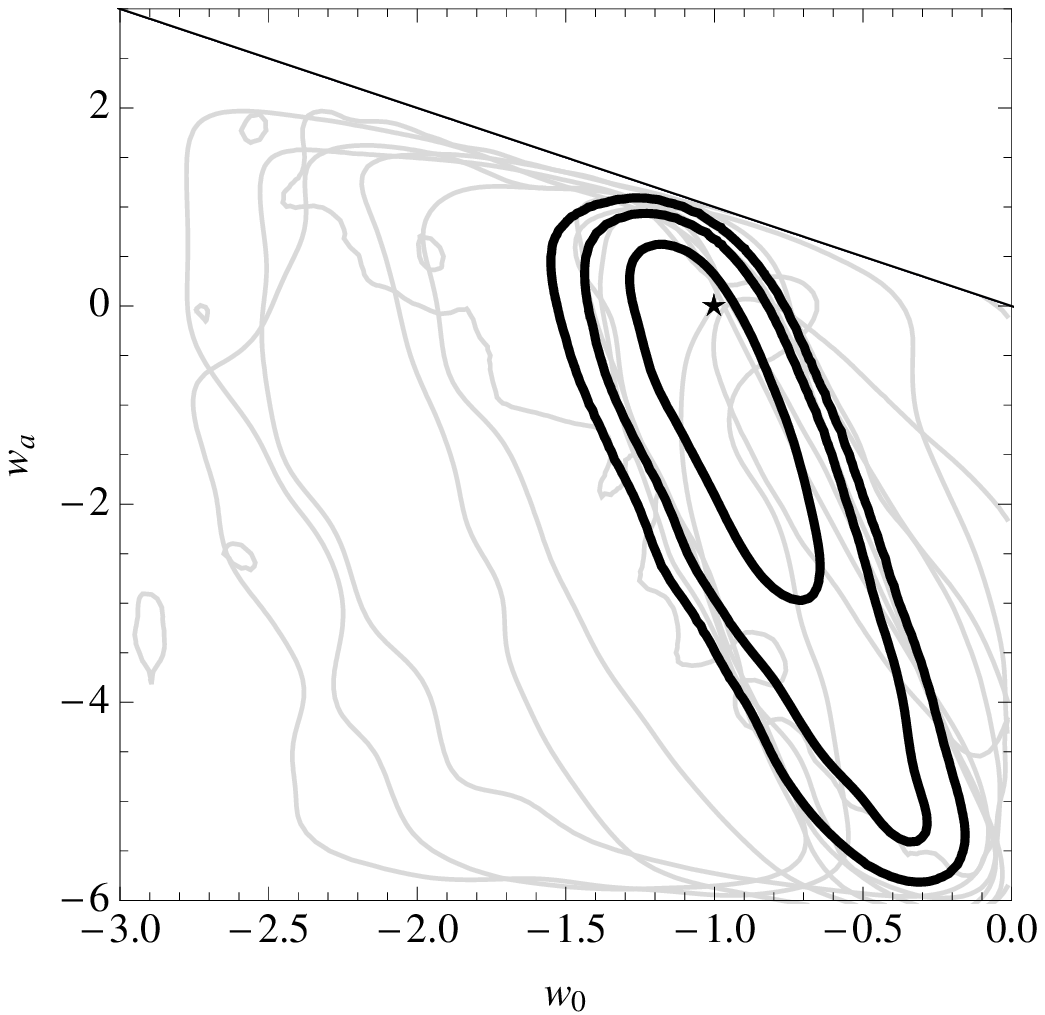}} 
\end{center}
\caption{Simulated CSL constraints on the dark energy
equation-of-state derived from $10$ clusters with $20$
multiply-imaged families each.  The dark contours show the $68$, $95$, and $99$ per
cent confidence regions in the ensemble result.  The light contours correspond to the 1-$\sigma$ regions from individual clusters.  The stars correspond to the input model
($\Lambda$CDM).  The simulated image catalogs include perturbations
due to scatter in the cluster galaxy populations, line-of-sight halos,
and observational errors on the image positions and redshifts. Top
panel: marginalized PDF in the $w_x$ and $\Omega_m$ plane.  Bottom
panel: simulated constraints on the CPL parameterization for the same
ten clusters as the top panel in the $w_a - w_0$ plane.  The line shows the border of the $w_a < -w_0$ prior.  We note that 
despite the inability to reproduce observed image positions to within $\sim 1$
arcsecond (see Table \ref{Tab:ClusterProperties}), useful constraints
can be obtained by combining results from independent clusters.}
 \label{LCDMcomb10}
\end{figure}

We first ray trace the sources through the LOS and cluster lensplanes
to obtain a catalog of image positions.  In order to obtain observed
positions, we must account for the fact that light from the cluster
itself is deflected by intervening matter.  We therefore lens the
cluster center through the lensplanes between the cluster and the
observer, and calculate the image positions with respect to this
center.  We also include the effect of lensing on cluster galaxy positions and magnitudes.  Finally,
image positions are randomly shifted by $\pm 2$ pixels ($0.05$
arcseconds per pixel) in order to roughly simulate space-based
observational errors.  We also redraw the source redshifts from normal
distributions with variance equal to 0.001 to simulate measurement
errors in spectroscopic redshifts. The final catalog of image
positions and redshifts of strongly lensed sources constitutes the
only set of constraints used in parameter recovery.

We estimate the total errors for each constraint using the Monte-Carlo
simulations in the last section.  Since the Monte-Carlo realizations
tend to be distributed along the local principal magnification axis,
for each image we transform to the coordinate system in which the
magnification tensor is diagonal.  In this coordinate system, the
distribution is typically well approximated by a bi-variate normal
distribution with zero correlations. Note that, in addition to the
errors, the numerator of each $\chi^2$ contribution is calculated in
the transformed coordinate system during parameter recovery.  Modeling
errors due to the LOS halos, scatter in cluster galaxy scaling
relations, and an observational error of 0.1 arcsecond are added in
quadrature for each image.  

Note that we have used the fiducial input models for each cluster to estimate errors.  In practice, one does not have a priori information about the smooth cluster component, cluster-galaxy scaling relations, or source positions.  This information would be required to perform the above Monte Carlo simulations.  An initial fit using the image positions would therefore be necessary as a starting point for the Monte Carlo simulations in this work.  Using the initial fit, one could then transform the observed images to the source plane.  Since the initial mass model would not be a perfect fit, images from the same source would not necessarily trace back to the same source position.  One way around this problem is to take the barycenters of positions corresponding to single sources.  The set of barycenters would constitute a model source catalog to base the Monte Carlo simulations on.  For more details on how Monte Carlo based error estimation would be performed in practice, we refer the reader to a recent paper on the cluster Abell 1689 that we have co-authored: \citet{Jullo2010Sci}.

We first illustrate the importance of correctly estimating
uncertainties.  Figure~\ref{ImageRedshiftPL} shows the contours of the
marginalized PDF obtained from a single cluster using a mock image
catalog of $20$ families which has been perturbed by scatter in the
cluster galaxies, LOS halos, and observational errors.  From here on,
the outer, middle and inner contours of PDFs correspond to 99, 95, and
68 per cent levels respectively.  We utilize the simple model
described in Section \ref{SEC:methods}, whose parameters are listed in
Table \ref{ParameterTable} to perform the parameter recovery.  The
dashed contours show the confidence regions obtained using the full
error estimates, which account for all perturbations
to image locations in our simulations.  The solid contours in the top
panel underestimate the uncertainties by assuming only observational
errors of $0.1$ arcsecond.  Note that this is an order of magnitude
smaller than typical perturbations obtained in Section
\ref{SEC:modelingerrors}.  The input model, denoted by a star, is
definitively ruled out when the errors are underestimated. The bottom panel
of Figure~\ref{ImageRedshiftPL} shows the effect of using photometric
redshifts.  We simulate errors for the high-redshift half of the image
catalog by redrawing from Gaussian distributions with standard
deviations of 0.5.  The full positional errors are used for both sets
of contours in the bottom panel.  Spectroscopically measured redshifts
are necessary to minimize the effect of observational uncertainties in
CSL cosmography.

Figure~\ref{LCDMcomb10} shows the results of combining constraints
from our simulated ten-cluster sample, where all errors have been
fully accounted for.  The top and bottom panels correspond to the
constant equation-of-state and CPL parameterizations respectively.  We checked the stability of the results by running the MCMC ten times for each cluster.  Table \ref{Tab:ClusterProperties} shows the properties of each cluster
in the sample.  The image plane root-mean-square (RMS) deviations are
frequently used in strong lensing studies to quantify how well a
particular model reproduces the observed images. For a given family
with $n$ images it is defined to be

\begin{equation}
\mathrm{RMS} = \sqrt{\frac{1}{n} \sum_{i=1}^{n}{\left( \vec{\theta}^i_o - \vec{\theta}^i \right)^2}}
\label{Eq:RMS}
\end{equation}
where $\vec{\theta}^i_o$ is the observed image position and
$\vec{\theta}^i$ is the model image position.  The total RMS for a
given model is obtained by averaging (\ref{Eq:RMS}) over all observed
families.  The last column of Table \ref{Tab:ClusterProperties} shows
that, for a given cluster, the parametric models are not able to
reproduce image configurations to within $\sim 1$ arcsecond.  In fact,
we should not expect them to perform any better due to the errors
discussed in Section \ref{SEC:modelingerrors}.  However, Figure
\ref{LCDMcomb10} indicates that the effect of these errors can be
alleviated by using a relatively small sample of clusters. Dark energy
constraints from CSL can therefore start to be competitive with
constraints from other cosmological techniques upon combining just
$\sim10$ clusters with $\sim20$ families each.

%%%%%%%%%%%%%%%%TABLE 2
\ctable[ caption = Properties of simulated clusters.]{c  c  c  c }{

\label{Tab:ClusterProperties}

\tnote[a]{The total mass within a circular aperture of 1 Mpc.}

\tnote[b]{Averaged over all MCMC samples.  Here, we quote the results for the constant $w_x$ case.}

}{
\hline\hline
 Redshift & Mass\tmark[a] & Mass in  & Image plane\\
   & ($\times10^{15}~\Msun$) & core galaxies & RMS\tmark[b] \\ 
   & & ($\times10^{13}~\Msun$) & (arcseconds) \\
\hline
\\ [-1ex]
 0.208 & 1.19 & 7.4 & 1.01 \\  \\                 
 0.336 & 1.27 & 4.8 & 0.74 \\ \\
 0.184 & 1.11 & 4.6 & 0.76 \\ \\
0.246 & 1.41 & 4.3 & 1.02 \\ \\
0.241 & 1.37 & 1.4 & 0.71 \\  \\
0.278 & 1.38 & 3.2 & 0.52 \\  \\
0.284 & 1.16 & 3.3 & 0.98 \\  \\
0.226 & 1.34 & 6.1 & 0.49 \\  \\
0.204 & 1.39 & 4.0 & 0.70 \\  \\
0.244  & 1.39 & 7.0 & 1.30 \\              
[1.5ex]
\hline
}
%\end{tabular}
%\label{ParameterTable}
%\end{ctable}

%%%%%%%%%%%%%%%%%%%%%%%%%%%

\section{Cluster mass profiles}
\label{SEC:clustermass}   

The main goal of this section is to show that CSL cosmography is not
limited to a particular type of cluster mass profile.  As pointed
out in Section \ref{SEC:imagelocations}, the CSL technique is a
geometric probe that exploits the dependence of lensed image locations on
angular diameter distances. In principle, it can be applied to any
strong lensing system, irrespective of the mass distribution of the
lens, as long as a sufficient number of constraints are observed to accurately
 model the system.

Although we have used PIEMD lenses exclusively in this work, our results
are not contingent upon this choice of profile.  To illustrate
this point, we use the Navarro-Frenk-White profile, which has a cusp
with the density approaching $r^{-1}$ in the inner regions, and
transits to $r^{-3}$ in the outer regions beyond the scale radius
\citep{1995MNRAS.275...56N,1996ApJ...462..563N,1997ApJ...490..493N}.
This profile is significantly different from the PIEMD case.  For lensing
properties of the NFW profile and an elliptical extension of it, we
refer the reader to \citet{1996A&A...313..697B} and
\citet{2002A&A...390..821G} respectively.

\begin{figure}
\begin{center}
\resizebox{8.0cm}{!}{\includegraphics{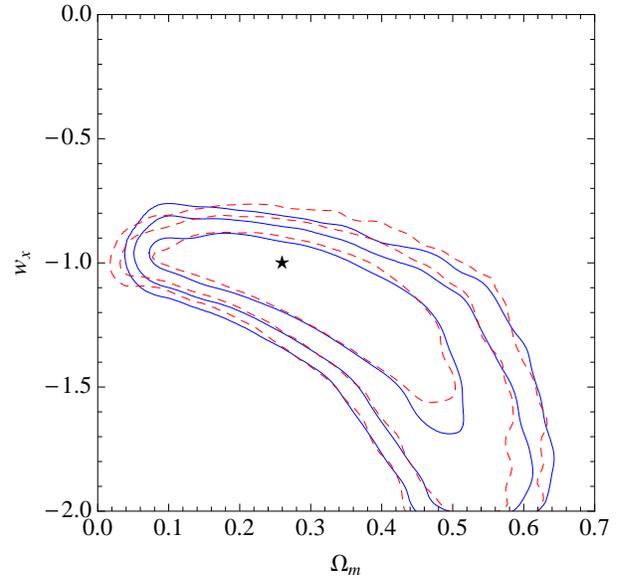}}
 \end{center}
\caption{An illustration that similar dark energy constraints may be obtained from clusters with very different mass profiles.  The solid and dashed contours correspond to PIEMD and NFW lenses respectively.  For both cases we use mock catalogs of 21
images. We assume only observational errors of 0.1 arcseconds for
each image.}
\label{FIG:NFWPIEMD}
\end{figure}

For comparison, we generate elliptical PIEMD and NFW lenses at a
redshift of $z_l = 0.21$ with equal masses of $1.6 \times
10^{15}~\Msun$.  The former has a velocity dispersion of 1300 km/s,
core radius of 41 $\kpc$ and scale radius of 900 $\kpc$.  The latter
has a concentration parameter $c = 4.5$ and a scale radius $r_s =
495\,\kpc$.  Both have ellipticity parameters of $0.3$. Since our
purpose in this section is solely to compare different cluster-scale
mass profiles, we neglect the role of sub-structure in what follows.
We lens a simulated source distribution for both cases and use the
same number of images as constraints for parameter recovery.  We run
the MCMC sampler using the input models with flat priors.  In the NFW
case all parameters are free.  In the PIEMD case, we fix $\rcore$ to
the input value so that the number of free parameters is the same in
the two cases.  Figure~\ref{FIG:NFWPIEMD} compares the results from
both cases.  The solid and dashed contours show the PIEMD and NFW
constraints respectively.  We note the similarity of the results,
indicating that a sample of NFW cluster lenses would generally yield
similar constraints to those shown in Section
\ref{SEC:simulatedconstraints}.
  
\begin{figure}
\begin{center}
\resizebox{8.0cm}{!}{\includegraphics{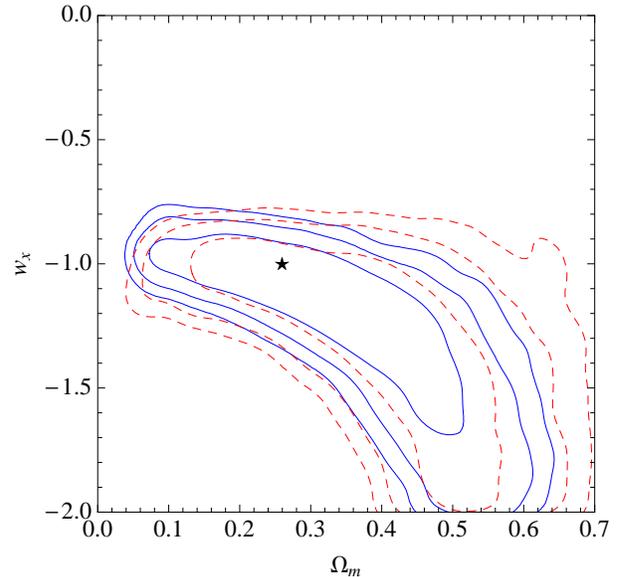}}
 \end{center}
\caption{Same as in Figure~\ref{FIG:NFWPIEMD}, except that the dashed
contours correspond to a bimodal cluster consisting of two PIEMD mass
peaks. Despite having twice the number of free parameters, the
bimodal constraints are very similar to the unimodal case.}
\label{FIG:bimodal}
\end{figure}

In this work we have utilized unimodal lenses consisting of only one
large-scale halo and a population of galaxy-scale sub-halos to model the mass 
distribution.  In
reality, cluster lenses can be more complex than this, requiring a
multi-modal mass distribution to fit the observed images.  Here, we
address the question of whether bimodal clusters (for example, Abell
1689; Abell 2218; Cl0024+16) can be useful for CSL cosmography.

We use the simulated PIEMD cluster above as a unimodal reference.  We
simulate bimodal lenses with mass equally divided between the two mass
peaks and total masses equal to the reference lens.  We generate 10
different lensing configurations for the bimodal case using 10 unique
source catalogs and random orientations of the two mass peaks.  We
then recover the cosmological parameters for each realization using
the same number of images as the reference configuration. Results from
the bimodal realization with the widest constraints are shown as the
dashed contours in Figure~\ref{FIG:bimodal}.  The solid contours
correspond to the unimodal reference lens.  Surprisingly, despite the
increase in the number of free parameters, the bimodal constraints are
generally similar to the unimodal constraints.   This result is robust
for various bimodal configurations and for various ratios of the two mass
peaks.  The tightness of the bimodal constraints is likely due to both
the higher amplification of images, on average, and to degeneracies
between parameters which limit the number of effective free
parameters.  We conclude that despite the larger number of parameters
required to model bimodal cluster lenses, these systems may be just as
useful for CSL cosmography as their unimodal counterparts.  Note,
however, that we have considered a highly idealized model and have not
taken into account the increased modeling errors that would likely
result from more complex mass distributions. In addition to the
difficulty in modeling more complicated systems, the increased errors
would act to weaken the dark energy constraints and possibly introduce
biases.  We defer a more detailed analysis of more realistic cluster
mass distributions derived from high resolution cosmological N-body simulations 
to future work.

\section{Discussion}
\label{SEC:discussion}

We have shown that, using a relatively small ($\sim 10$ galaxy clusters)
sample of well constrained cluster lenses with parametric mass models,
it may be possible to obtain constraints on the dark energy equation
of state and its time variation that are competitive with existing
probes.  Cluster strong lensing may provide a very useful
complementary tool, particularly in probing dynamic dark energy
models, which are currently poorly constrained, in the near future.

We have used simple Monte-Carlo simulations of cluster lensing
configurations to explore some of the potentially largest sources of
modeling errors.  Owing to the large number of parameters that would
be required to accurately model the cluster galaxy population,
parametric models must assume scaling relations between galaxy
luminosities and their mass profile parameters.  While these
assumptions generally improve agreement with observed images, we have
shown that scatter in the scaling relations can introduce modeling
errors as large as $\sim 1$ arcsecond for clusters at $z \sim
0.2-0.3$.

We have used publicly available halo catalogs from the Millennium
Simulation in order to quantify the lensing effects of intervening
halos along the line-of-sight between the observer and the background
sources. Owing to limited observational resources at the present, most
of these halos would not be modeled in practice, and could potentially
add significantly to the CSL error budget. A redshift survey behind
the most massive lensing clusters would be required to explicitly include structures along the line of sight in the
modeling. We created an ensemble of line-of-sight realizations
containing analytic galaxy-scale potentials.  Using a multi-plane
lensing algorithm, we traced light rays through each line-of-sight
realization, with the simulated cluster placed at the appropriate
redshift.  Although LOS halos typically contribute less than $\sim10$
per cent to the efficiency-weighted surface mass densities, they can
introduce deflections with respect to cluster-only models that are
typically on the order of a few arcseconds for clusters at $z\sim
0.2-0.3$.  In rare cases, deflections can be as large as $\sim 10$
arcseconds.

Given the fact that scatter in the cluster galaxy population and LOS
halos perturb images typically by a few arc seconds, simple parametric
models cannot be expected to reproduce image locations to within this
accuracy.  We have shown that underestimating the total errors can lead to
severe biases in dark energy parameter recovery.  This is particularly
relevant in the case of space-based imaging and spectroscopically
obtained redshifts, where observational errors are typically of the
order of $\sim 0.1$ arcseconds - an order of magnitude lower than
potential modeling errors.

We have used our Monte-Carlo simulations to perform a feasibility test on
obtaining competitive dark energy constraints from CSL systems.  We
used a sample of $10$ clusters with $20$ multiply imaged sources each.
These sources have redshifts between $z\sim 0.7 - 5$.  The simulated
image catalogs take into account deflections due to scatter in the
cluster galaxy scaling relations and intervening LOS halos.  We also
simulated observational errors on positions and redshifts.  For each
simulated cluster lens, we used a Bayesian MCMC technique to probe the dark energy equation-of-state.  We found that the observed images only
need to be reproduced to within $\sim1$ arcsecond on average to get
useful constraints upon combining results from $\sim 10$ clusters.

Our feasibility test expands on the work of \citet{2005ApJ...622...99D} in a number of ways.  Perhaps most importantly, we have shown that Bayesian Inference with Markov Chain Monte-Carlo can be used to avoid biases that may occur in a best-fit approach by probing the full range of parameters that are statistically compatible with the image data.  Crucial to this point is the use of accurate error estimates on image positions, obtained by a detailed analysis of each individual cluster, and models containing sufficient degrees of freedom to incorporate complexities such as substructure.  The large range of cosmological parameters compatible with a single lensing system can be narrowed by combining results from different systems.    

%and Bayesian Markov Chain Monte-Carlo  each cluster model used for the parameter inference in our test includes a simplified component to model the influence of substructure.  In section 3 of their paper, .In practice, the modeling of substructure is crucial to avoiding biased constraints  We have also quantified modeling errors for each cluster and included them in the parameter inference.  In practice, error estimates will be performed 

%Finally, an important aspect of our approach was the use Bayesian Inference with Markov Chain Monte-Carlo.  

%for each \citet{2005ApJ...622...99D} point out utilized.  We also illustrated the effect of combining constraints from individual clusters.

While our simulations have allowed us to explore some of
the main difficulties in CSL cosmography in a straightforward way,
there are some important limitations which we plan to address in
future work.  First, our calculation on the effects of the
line-of-sight includes only mass in collapsed halos within the field of view.  In reality,
the light rays are influenced by an extended network of larger scale structures.  The
mass exterior to halos also contributes significantly to the overall deflection of light
rays.  However, for our purposes, the relevant quantities are the relative deflections between the
observed images, which are typically separated by tens of arcseconds.
We plan to explore the impact of extended large-scale structure in future work exploiting the full particle data and periodic boundary conditions of a high resolution cosmological N-body simulation.

Another key limitation is that we have used the same parametric profile for the smooth, large-scale component of both the input and recovery models.  While our focus in this work was primarily on the influence of substructure and line-of-sight halos, it will be important in future work to test the influence of uncertainties in the large-scale component.  In Figure 2 of their paper, \citet{2005ApJ...622...99D} show that using a parametric profile different from the input model can lead to significant biases in the parameter recovery.  We have performed our own tests attempting to recover PIEMD lenses with NFW models and vice versa.  We found that the Bayesian sampler did not converge and that the poorness of fit was reflected in the evidence values.  In practice, it should be straightforward to determine if the model for the large-scale component generally fails to reflect the true mass distribution (i.e. utilizing RMS deviations between model and observed images and comparison of Bayesian evidence values).  However, uncertainties due to less obvious deviations in the large-scale component is a topic that needs to be explored.

In upcoming work, it will also be crucial to test
how well parametric models can be used to describe more complex
cluster mass distributions.  In principal, more sophisticated models can be introduced for these systems \citep[e.g.][]{2005MNRAS.360..477D,2005MNRAS.362.1247D,2008ApJ...681..814C,2009MNRAS.395.1319J,2010arXiv1005.0398C}.  With
$\sim20$ families, there are certainly enough constraints to increase
the number of free parameters. However, doing so may degrade the efficacy
of derived cosmological constraints and introduce additional modeling
errors and degeneracies.  It may be possible to alleviate this problem
by simply obtaining a larger sample of cluster lenses or including
additional independent constraints. Note that we have investigated the
use of strongly lensed image positions and redshifts, but have not
considered other possible sources of information.  Inclusion of X-ray
gas temperature profiles, measured cluster velocity dispersions, and
the stellar velocity dispersion profiles of bright cluster galaxies
may enable the tightening of derived cosmological constraints with
more sophisticated mass models.

We conclude with a brief discussion on observational strategy.  Clearly, the ideal cluster lensing system for cosmography would have a relatively simple mass distribution with a large number of observed images.  However, in practice, there will likely be a trade-off between complex lensing systems that produce a larger number of strongly lensed images and simple systems with fewer constraints.   This is due to the possible enhancement of strong lensing cross sections from substructure and asymmetries \citep[e.g.][]{1995A&A...297....1B,2007A&A...461...25M}, mergers \citep[e.g.][]{2004MNRAS.349..476T,2006A&A...447..419F} and line-of-sight matter \citep[e.g.][]{2005ApJ...635L...1W,2007ApJ...654..714H,2007MNRAS.382..121H,2009MNRAS.398.1298P}.  The most dramatic and readily observable cluster strong lenses are likely to be more complicated to model.  However, in section 6 we showed that a larger number of model parameters does not necessarily correspond to weaker dark energy constraints.  Therefore, it may not be wise to devote all resources towards the simplest lensing systems with the fewest sources of uncertainty.  The complex systems are probably easier to find and may provide equal or more leverage on cosmological parameters.  These benefits may outweigh the costs of more detailed analyses.  Complex and simple cluster lensing systems are likely to play complementary roles in cosmographic applications.

\section*{Acknowledgments}
The authors thank Eric Jullo and Jean-Paul Kneib for insightful discussions and the anonymous referee for helpful comments and suggestions.  PN acknowledges the receipt of a Guggenheim Fellowship from the John P. Simon Guggenheim Foundation, a Radcliffe Fellowship from the Radcliffe Institute for Advanced Study, and a grant from the National Science Foundation's Theory Program (AST10-44455).       

\bibliographystyle{mn2e} 
\bibliography{CSL}

\end{document}